\documentclass[10pt,letterpaper]{article}
\usepackage[top=0.85in,left=2.75in,footskip=0.75in,marginparwidth=2in]{geometry}

\usepackage[utf8]{inputenc}

\usepackage{cite}

\usepackage{nameref,hyperref}

\usepackage{microtype}
\DisableLigatures[f]{encoding = *, family = * }

\raggedright
\usepackage{changepage}

\usepackage[aboveskip=1pt,labelfont=bf,labelsep=period,singlelinecheck=off]{caption}

\makeatletter
\renewcommand{\@biblabel}[1]{\quad#1.}
\makeatother

\usepackage{lastpage,fancyhdr,graphicx}
\usepackage{epstopdf}
\fancyheadoffset[L]{2.25in}
\fancyfootoffset[L]{2.25in}

\usepackage{color}

\definecolor{Gray}{gray}{.25}

\usepackage{graphicx}

\usepackage{sidecap}
\usepackage{booktabs}
\usepackage{wrapfig}
\usepackage[pscoord]{eso-pic}
\usepackage[fulladjust]{marginnote}
\reversemarginpar

\begin{document}
\vspace*{0.35in}

\begin{flushleft}
{\Large
\textbf\newline{Research on the knee region of cosmic ray by using a novel type of electron-neutron detector array}
}
\newline
\\
Bing-Bing Li\textsuperscript{1},
Xin-Hua Ma\textsuperscript{2,3,*},
Shu-Wang Cui\textsuperscript{1,*},
Hao-Kun Chen\textsuperscript{4,5},
Tian-Lu Chen\textsuperscript{4,5},
Danzengluobu\textsuperscript{4,5},
Wei Gao\textsuperscript{2,3}
Hai-Bing Hu\textsuperscript{4,5}
Denis Kuleshov\textsuperscript{6}
Kirill Kurinov\textsuperscript{6}
Hu Liu\textsuperscript{7}
Mao-Yuan Liu\textsuperscript{4,5}
Ye Liu\textsuperscript{8}
Da-Yu Peng\textsuperscript{4,5}
Yao-Hui Qi\textsuperscript{1}
Oleg Shchegolev\textsuperscript{6,9}
Yuri Stenkin\textsuperscript{6,9}
Li-Qiao Yin\textsuperscript{2,3}
Heng-Yu Zhang\textsuperscript{4,5}
Liang-Wei Zhang\textsuperscript{1}
\\
\bigskip
\bf{1} College of Physics, Hebei Normal University, Shijiazhang 050024, China
\\
\bf{2} Key Laboratory of Particle Astrophysics, Institute of High Energy Physics, Chinese Academy of Sciences, Beijing 100049, China
\\
\bf{3} TIANFU Cosmic Ray Research Center, Chengdu, Sichuan 610000, China
\\
\bf{4} Science School, Tibet University, Lhasa, Tibet 850000, China
\\
\bf{5} Key Laboratory of Comic Rays, Ministry of Education, Tibet University, Lhasa, Tibet 850000, China
\\
\bf{6} Institute for Nuclear Research of the Russian Academy of Sciences, Moscow 117312, Russia
\\
\bf{7} School of Physical Science and Technology, Southwest Jiaotong University, Chengdu, Sichuan 610031, China
\\
\bf{8} School of Management Science and Engineering, Hebei University of Economics and Business, Shijiazhuang, Hebei 050061, China
\\
\bf{9}Moscow Institute of Physics and Technology, 141700 Moscow, Russia
\bigskip

* maxh@ihep.ac.cn, cuisw@hebtu.edu.cn

\end{flushleft}

\section*{Abstract}
By accurately measuring composition and energy spectrum of cosmic ray, the origin problem of so called ``knee" region (energy $>$ 1 PeV) can be solved. However, up to the present, the results of the spectrum in the knee region obtained by several previous experiments have shown obvious differences, so they cannot give effective evidence for judging the theoretical models on the origin of the knee. Recently, the Large High Altitude Air Shower Observatory (LHAASO) has reported several major breakthroughs and important results in astro-particle physics field. Relying on its advantages of wide-sky survey, high altitude location and large area detector arrays, the research content of LHAASO experiment mainly includes ultra high-energy gamma-ray astronomy, measurement of cosmic ray spectra in the knee region, searching for dark matter and new phenomena
 of particle physics at higher energy. The electron and Thermal Neutron detector (EN-Detector) is a new scintillator detector which applies thermal neutron detection technology to measure cosmic ray extensive air shower (EAS). This technology is an extension of LHAASO. The EN-Detector Array (ENDA) can highly efficiently measure thermal neutrons generated by secondary hadrons so called ``skeleton" of EAS. In this paper, we perform the optimization of ENDA configuration, and obtain expectations on the ENDA results, including thermal neutron distribution, trigger efficiency and capability of cosmic ray composition separation. The obtained real data results are consistent with those by the  Monte Carlo simulation.
%

\section{The knee region in the cosmic ray spectrum}
The study on the cosmic rays has remained a relatively young discipline since it is initiated by Victor Hess's famous discovery over 110 years ago. It still needs to establish a comprehensive ``standard model" on describing their origin, acceleration mechanisms, and the way of propagation through interstellar space, as well as the modulation effects induced by solar activities upon their entering into the solar system\cite{Solar, Mode}.
Cosmic ray energy spectrum spans over many orders from about 10$^{6}$ eV to beyond 10$^{20}$ eV, with its flux showing the power law with different indices and decreasing along primary energy. At the energy levels of about 10$^{15}$ eV, there exists a so called ``knee" region, with the power law index changing from approximately -2.7 to -3.0. Various theoretical models have been tested to study this phenomenon based on the extensive experimental data accumulated, and meticulous analyses and comparisons, but it is still limited by the insufficient knowledge regarding the sources of cosmic ray. Therefore, tracing the origins of cosmic ray becomes crucial in the cosmic ray study, which is challenging because cosmic ray travels through the pervasive magnetic fields existed throughout the universe over millions of years and charged cosmic ray may have lost all the information about their origins. We have to capture the neutral particles, namely photons and neutrinos, generated within and thus carrying information about their original acceleration sources. This is the motivation for the establishment of what is called the ``Ultra High Energy (UHE, $E_{\gamma}$ \textgreater 0.1 PeV) $\gamma$-ray observation".

The Large High Altitude Air Shower Observatory (LHAASO)~\cite{LHAASO} is a complex of the Extensive Air Shower (EAS) detector arrays, located at Mt. Haizi ($29^\circ21'27.56''$ N, $100^\circ 08'19.66''$ E, 4410 m a.s.l.) in Daocheng, Sichuan province, P.R. China. In 2021, LHAASO has reported a significant amount of UHE photons from 12 cosmic $\gamma$-ray sources. This reveals that our galaxy is actually full of PeVatrons, almost evenly distributed in the disk. Furthermore, two super-PeV photons have been detected from the Crab at 1.1 PeV and the Cygnus region at 1.4 PeV~\cite{PeVatrons, Crab}. 
This indicates the existence of super-PeVatrons in our galaxy. The super-PeVatrons, certainly not limited to those two, may also be responsible for the flux of cosmic ray above the knee region. These discoveries mark the onset of the UHE $\gamma$-ray astronomy~\cite{Cao}. On 9 October 2022,  LHAASO observed GRB221009A, the brightest gamma-ray burst ever recorded ~\cite{Fermi, BAT, GRB}. For the first time, LHAASO documented the entire process of the TeV $\gamma$-ray flux enhancement and decay. On the other hand, By accurate measurement of composition and energy spectrum, the origin problem of the cosmic ray knee region can be solved by  giving clear evidence in judging  the various theoretical models~\cite{keen-asp1, keen-asp2, keen-nucl1}. With the rapid development of the ground-based EAS experiments, represented by LHAASO, we are getting closer to unlocking the secret of the knee region in cosmic ray spectrum.

\section{The status of the knee region in cosmic ray spectrum and the novel detecting technology}
The ``knee problem" in the cosmic ray study has existed for long, over 65 years. The most popular explanation on the observed power law break in the EAS size was assigned to astrophysical reasons, arguing that the knee is an intrinsic property of the energy spectrum~\cite{keen-asp1}, i.e. to the primary cosmic ray spectrum and its propagation in the cosmic media. The appearance of the knee represents the limit of the spectrum of cosmic ray accelerated by supernova remnants in the Milky Way~\cite{keen-asp2}. But the EAS method reason is only a secondary one and other non-astrophysical reasons can also be responsible for appearance of the knee. Some published works propose alternative explanations on the knee phenomenon, including from the nuclear physical aspect~\cite{keen-nucl1, keen-nucl2, keen-nucl3} by assuming the spectrum being changed in the hadronic interactions, and/or due to the generation of new or unknown particles, etc., and from the phenomenological side~\cite{keen-nucl4} to explain the knee in the PeV region by assuming the change in the EAS structure at the $\sim$100 TeV/nucleon when the EAS cascading hadrons reach the observation level. This means that below the hadron energy the EAS will be hadronless and coreless and its properties will differ significantly from that of a normal EAS where there exists an equilibrium between the EAS components. It is observed that the equilibrium is reached at the energy when the heaviest primary species (the iron group) exceed the threshold and when the EAS method starts to work correctly, i.e. at E \textgreater 6 PeV or so. To exclude the hadronless showers automatically, the hadronic component has to be measured over the full detector array area.

The knee region was measured in several experiments, e.g., KASCADE~\cite{KASCADE}, Tibet-AS$\gamma$~\cite{Tibetasg}, ARGO/WFCTA~\cite{argo-wfcta}, etc. Although they have confirmed the existence of the knee region, these experiments exhibited different results either with systematic deviations, or in terms of the exact energy location of the knee, in the magnitude of the flux before and after the knee, in the values of power law indices, or in the proportions of the cosmic ray composition. Such results failed to provide clear evidence for making effective judgement on the various theoretical models proposed to explain the origin of the knee in cosmic ray spectrum~\cite{keen-asp1, keen-asp2, keen-nucl1}.

Up to date, no experiments, including LHAASO, have measured hadrons in EAS, which possess the most energetic secondary particles and define the ``skeleton" of the EAS via hadronic cascade. They may also carry important information required for the multi-parameter correlation studies. The hadrons detection was limited for a long time by the absence of simple and cheap hadron detectors. Conventional hadron calorimeters (HCAL) is expensive, complicated, and hard to deploy over large areas. Then a novel way to deal with this problem was brought about, i.e. using nuclear reactions between the hadrons and matters in the surrounding environment (such as soil, buildings, detector materials, air) to produce a large amount of evaporation neutrons of up to two orders of magnitude more than the parent hadrons~\cite{StenkinMPLA, StenkinPRISMA, StenkinCPC}. The evaporation neutrons are then moderated to thermal neutrons by the matters in the surrounding environment. Based on the principle of detecting thermal neutrons, the Primary Spectrum Measurement Array (PRISMA) project was proposed and led to the design of the electron and thermal neutron detector (EN-detector) to measure both thermal neutrons and charged particles~\cite{StenkinPRISMA, StenkinPAN, StenkinNPPS2008}.

\begin{figure}[ht!]
\centering
\includegraphics[width=.9\textwidth]{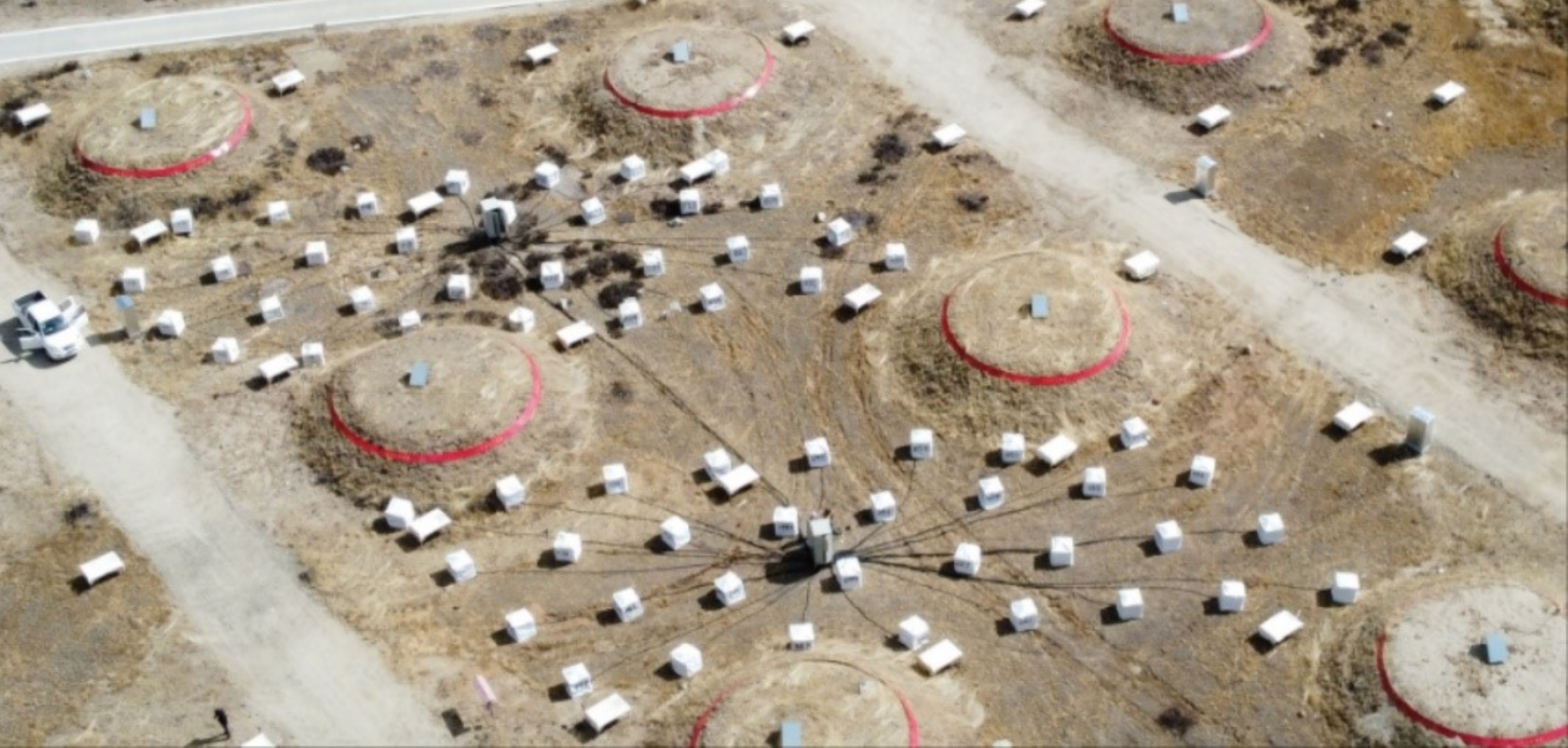}
\caption{Photo of the ENDA-64 detector arrays in a bird view.}
\label{ENDA-photo}
\end{figure}

The idea was implemented at high altitude: one small array called PRISMA-YBJ was installed inside a hall hosting the ARGO-YBJ experiment at the Yangbajing (YBJ) Cosmic Ray Observatory (Tibet, China, at 4300 m above sea level)~\cite{ARGO-PRISMA}. PRISMA-YBJ was composed of four EN-detectors based on an inorganic scintillator ZnS(Ag) added with LiF enriched with the $^6$Li isotope up to 90\%. Between PRISMA-YBJ and ARGO-YBJ, the coincident EAS events generated by cosmic rays were obtained, and a positive correlation between the thermal neutrons and the electromagnetic component generated in EAS was confirmed. Besides, it was indicated that the EN-detectors can also be used to monitor seismic activities~\cite{StenkinPAG, StenkinJER}. Following PRISMA-YBJ, the Electron-Neutron Detector Array (ENDA) was decided to deploy in the LHAASO~\cite{LHAASO} and the pro-phase works were performed. In which, a new type of EN-detectors was developed with the neutron capture isotope $^{6}$Li replaced by $^{10}$B, and two prototype arrays each with 16 EN-detectors were built: PRISMA-YBJ-16 at YBJ~\cite{ENDA-JINST2017}\cite{ENDA-ASS2020}, and ENDA-16~\cite{ENDA-PAN2021} at LHAASO. Several studies for testing the array performance were made: the negative correlation between the thermal neutron counting rate and soil moisture was obtained~\cite{ENDA-ASS2020, ENDA-JINST2023}, and the so-called ``sand cubes" were installed to study the influence of the target material, which is a major environmental factor affecting the arrays~\cite{ENDA-ASS2022}. In March 2023, ENDA-64 composed by 64 EN-detectors has started running focusing on the knee region of the light component (Fig.~\ref{ENDA-photo}).

\section{The electron-neutron detectors array}
\subsection{Principle of the detection}
Among isotopes used in thermal neutron capture, $^{6}$Li releases the highest energy for detection during the reaction, but it is a rare material generated during nuclear fission and is strongly limited for purchase with quite high price. Although $^{10}$B capturing neutron with lower released energy than $^{6}$Li, it has a relatively larger cross section. Moreover, natural Boron contains 19\% of $^{10}$B versus only 7\% of $^{6}$Li in natural Lithium. After comprehensive considerations, we choose to replace $^{6}$Li by $^{10}$B when building ENDA. One $^{10}$B captures a thermal neutron via the following nuclear reaction,

\begin{equation}
\left. \begin{array} {l}
^{10}B + n \rightarrow  ^{7}Li^{\star} + \alpha \\
\rightarrow  ^{7}Li + \alpha + 0.48 MeV \gamma + 2.3 MeV (93\%)  \\
\rightarrow ^{7}Li + \alpha + 2.8 MeV (27\%)
\end{array} \right.
\label{nB10}
\end{equation}

\begin{figure}[ht!]  
	\centering  
	\includegraphics[width=0.4\linewidth]{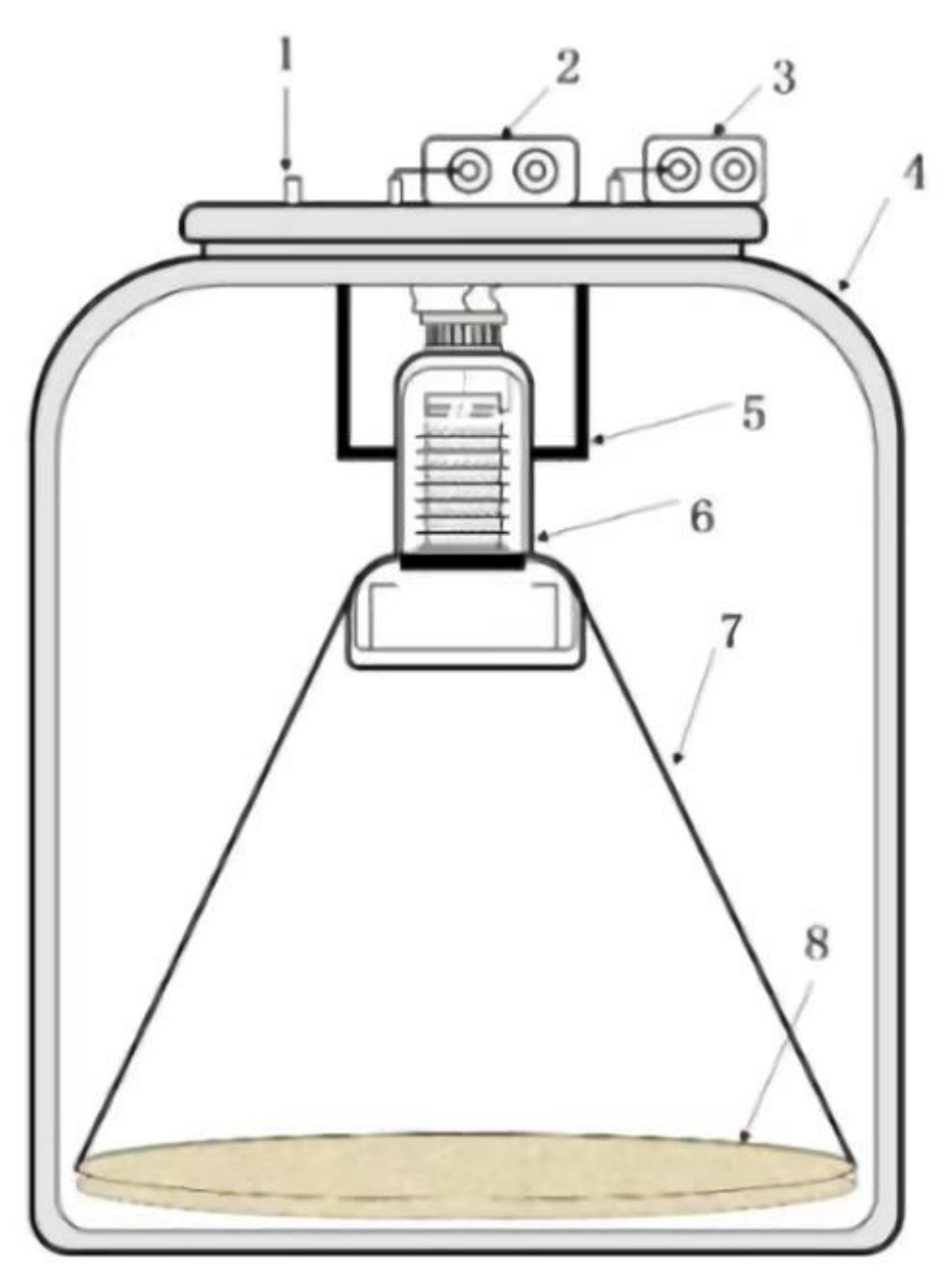}  
	\\[1mm]  
	\caption{
          Schematic diagram of the EN-detector. 1 – the high-voltage input port, 2 – DIU connected to the 8th dynode of the PMT, 3 – IU connected to the 5th dynode of the PMT, 4 - a black tank for the detector housing, 5 – the PMT fixed holder, 6 – the PMT, 7 – the scintillation light collecting cone, and 8 – the scintillator.
	}\label{fig:Det}
\end{figure}

A novel type of ZnS(Ag) scintillator alloyed with B$_{2}$O$_{3}$ (65\% and 35\%, respectively) incuding the $^{10}$B isotope of about 19\% is developed. The powder of the ZnS(Ag) and B$_{2}$O$_{3}$ alloy is deposited in optical silicon rubber. The effective thickness of the scintillator layer is 50 mg/cm$^2$. The design of a typical EN-detector is shown in Fig.~\ref{fig:Det}. The scintillator is then placed at the bottom of a cylindrical polyethylene (PE) tank used as the detector housing. A 4-inch photomultiplier (PMT) (manufactured by Beijing Hamamatsu, model CR-165) is mounted on the tank lid, 0.3-m away from the scintillator. A conical reflective layer between the scintillator and the PMT is used to collect the scintillation photons. The thermal neutron detection efficiency of the scintillator layer is about 20$\%$. The front-end electronics (FEE) of each detector consists of a PMT voltage divider, a discriminator-integrator unit (DIU) and an integrator unit (IU). A DIU, including charge sensitive preamplifier, discriminator, and integrator, connects to the 8th dynode of the PMT for energy deposit measurements, a coincidence selection and counter counting the number of secondary neutrons. A IU, including charge sensitive preamplifier and integrator, connects to the 5th dynode of the PMT for expanding its dynamic range. The output signals of the EN-detectors include both the weak but fast signals generated from the charged particles, and the high amplitude but slow and delayed signals generated from the captured thermal neutrons.

Every 16 EN-detectors compose of a cluster aligned in the form of a 4 $\times$ 4 array with an adjacent distance of 5 m. A data acquisition system (DAQ) to each cluster consists of a 32-channel flash analog-to-digital converter (FADC) connected to a PC via an optic fiber. The first 16 channels of the FADC are used for receiving the signals from the 8th dynodes of the PMTs and the rest 16 for the 5th dynodes. The first pulse produced mostly by the EAS charged particles is used as trigger signal and for energy deposit measurements, and the delayed neutron capture pulses are counted within a time gate of 20 ms to give the number of detected thermal neutrons. If any 2 of the 16 detectors coincide within the window of 1 $\mu$s (trigger type M1), the FADC performs a sampling. This is the first-level triggering. The online program then analyzes the received input signals and perform the second-level triggering as follows~\cite{ENDA-JINST2017},

\begin{enumerate}
    \item[(1)] At least 2 detectors produce a first-level triggering, with signals $\ge$ 10 minimum ionization particles ($m.i.p.$s);
    \item[(2)] The total energy deposit power measured is 125 $m.i.p.$s;
    \item[(3)] The total number of the thermal neutrons recorded is $\ge$ 3.
\end{enumerate}

If condition (1) is satisfied, the trigger type is M1, if (1) and (2) are satisfied, the trigger type is M2, if (1) and (3) are satisfied, the trigger type is M3, and if (1), (2), and (3) are satisfied, the trigger type is M4. In addition, the online program generates a software trigger signal every 5 minutes to count the thermal neutrons in 20 ms as the thermal neutron background (M0 events).

\subsection{The simulation setup}
\begin{figure}[ht!]  
	\vspace*{1mm}  
	\centering  
	\includegraphics[width=10cm,height=8cm]{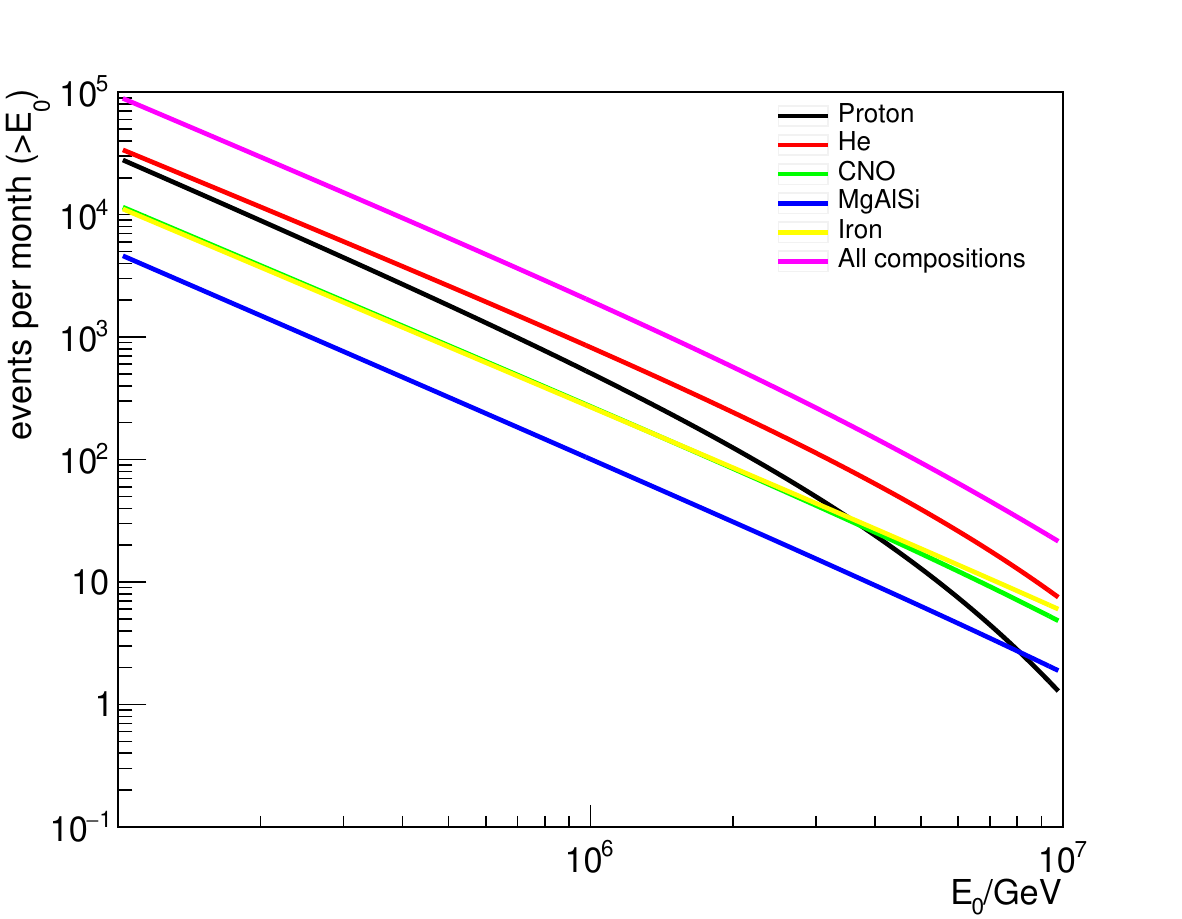}  
	\\[1mm]  
	\caption{
		Primary energy spectra of cosmic ray components normalized to the Gassier model.
	}\label{fig:energy}
\end{figure}

In the simulation, EAS of cosmic ray, including proton, He, CNO, MgAlSi, and Fe, are generated by CORSIKA~\cite{Corsika} version 7.640. The selected hadronic interaction models include QGSJETII for the high energy range and GHEISHA for the low energy range. The whole primary energy range is set from 100 TeV to 10 PeV. The zenith angle range is set from 0$^\circ$ to 40$^\circ$, the azimuthal angle range is set from 0$^\circ$ to 360$^\circ$, and the observation level is set at 4400 m above see level. The low energy cut-off of the secondary hadrons is set as 1 GeV, and the low energy cut-off of the secondary electromagnetic components and muons is set as 0.01 GeV. The energy spectra of all the cosmic ray components are normalized to the Gaisser H3a model~\cite{keen-asp2},

\begin{equation}
	J = \sum_{i=1}^{3}a_{i}\times E^{-\gamma_{i}}\times \mathrm{e}^{-\frac{E}{z\times Rc_{i}}} {\rm (GeV^{-1}m^{-2}sr^{-1}s{-1})}
\label{spectrum}
\end{equation}

where $a_i$ is a spectral constant,  $\gamma_{i}$ is the spectral index, $z$ is the atomic mass, $Rc_{i}$ is the rigidity of cosmic ray ( $Rc_{1}$= 4PV,  $Rc_{2}$= 30PV and $Rc_{3}$= 2EV ) (Tab.~\ref{tab1}). Considering the detection area by a cluster of the ENDA and the detection range in the zenith angle and in the azimuthal angle, a month integral spectra of all the components (Fig.~\ref{fig:energy}) show that one cluster of ENDA can obtain about 500 protons and 300 Fe events, respectively, above 1 PeV in one month.

\begin{table}
\centering
\caption{Parameters for the simulation function}
\begin{tabular}{ cc c c c c c c  }
	\midrule
	Composition & $a_1$ & $a_2$ & $a_3$ & $\gamma_1$ & $\gamma_2$ & $\gamma_3$\\
	\hline
	P            & 7860 & 20   & 1.7   & 2.66 & 2.44 & 2.44\\
	He         & 3550 & 20   & 1.7   & 2.58 & 2.44 & 2.44\\
	CNO     & 2200 &13.4 & 1.14 & 2.63 & 2.44 & 2.44\\
	MgAlSi & 1430 &13.4 & 1.14 & 2.67 & 2.44 & 2.44\\
	Fe          & 2120 &13.4 & 1.14 & 2.63 &2.44 & 2.44\\
	\midrule
\end{tabular}
\label{tab1}
\end{table}

The EN-detector response to secondary particles in EAS is simulated with GEANT4~\cite{geant4} version 10.4. The selected physical interaction model is QGSP\_BIC\_HP, i.e., a combination of the Quark Gluon String Model (QGS, \textgreater $\sim$ 20 GeV), the Binary Cascade Model (BIC, \textless $\sim$ 10 GeV), and the High Precision Neutron Model (HP, \textless 20 MeV). The core of each EAS event generated by the CORSIKA is randomly dropped with uniformity into an area of 20 m $\times$ 20 m around the center of the cluster. The dropping area is enough for our simulation because of the sharp decrease in the thermal neutrons along the lateral distribution which has been confirmed in the previous test~\cite{ARGO-PRISMA}. In order to increase the simulation speed, when a shower falls into the array, secondary particles in different types are selected according to the distance to the simulated detectors: one hadron falling within 4 m far from the detector, and one electromagnetic component ($\gamma$, e$^+$, e$^-$ , etc.) or one muon falling within area of the detector. The difference between the fast simulation and full simulation is 3\% for the secondary electromagnetic components and muons and 4\% for the secondary hadrons. The fast simulation can increase speed about 180 times for the secondary electromagnetic components and muons and 6 times for the secondary hadrons. Totally fast simulation can increase speed about 57 times.

\section{Results}
\subsection{Optimization of the array}
\begin{figure}[ht!]
	\vspace*{1mm}
	\centering
	\includegraphics[width=9cm,height=7.2cm]{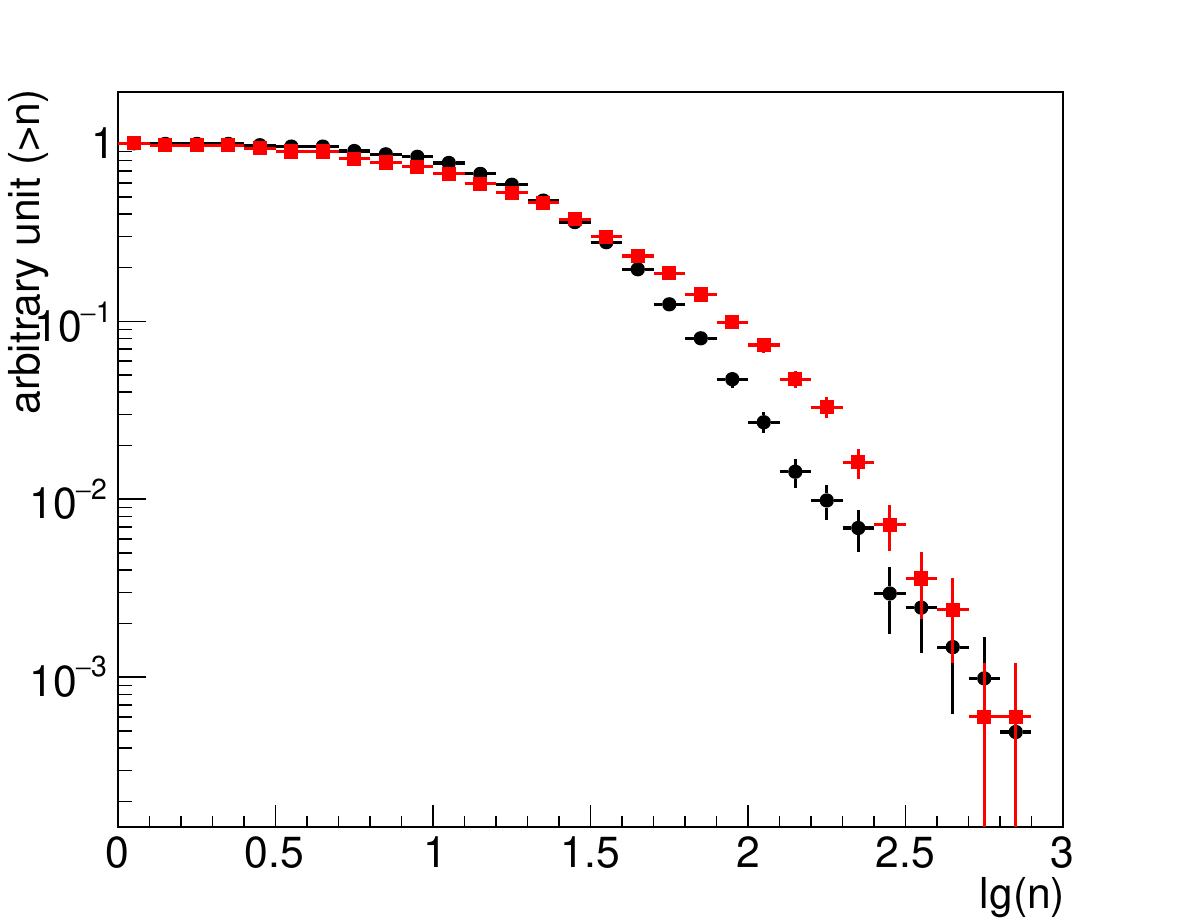}
	\\[1mm]
	\caption{
	The thermal neutron integrated distributions normalized to the number of the simulated events with different adjacent distance between the detectors. Black dots are for the case of 5-m distance, and red squares are for that of 3-m distance.
	}\label{fig:distance}
\end{figure}

In order to optimize the array alignment to achieve a higher performance-price ratio, the adjacent distance between the detectors is tuned to extend the array area to get higher acceptance while keeping the high detection performance for the cosmic ray energy spectrum measurement (Fig~\ref{fig:distance}). It indicates that the alignment with a 3-m adjacent distance between the detectors can collect more thermal neutrons than that with a 5-m distance. But the thermal neutron integrated distribution with the 3-m distance is not a single power law because the area of the detector array with the 3-m distance is too small. On the contrary, although collecting fewer thermal neutrons, the alignment with the 5-m distance keeps the detected thermal neutron integrated distribution in a single power law, which is necessary for the cosmic ray spectrum study. It confirms the previous test at ARGO-YBJ hall in which the lateral distribution of thermal neutrons was obtained by using 4 earlier thermal neutron detectors together with ARGO-YBJ RPC detectors array ~\cite{ARGO-PRISMA}.

\begin{figure}[ht!]
	\vspace*{1mm}
	\centering
	\includegraphics[width=9cm,height=7.2cm]{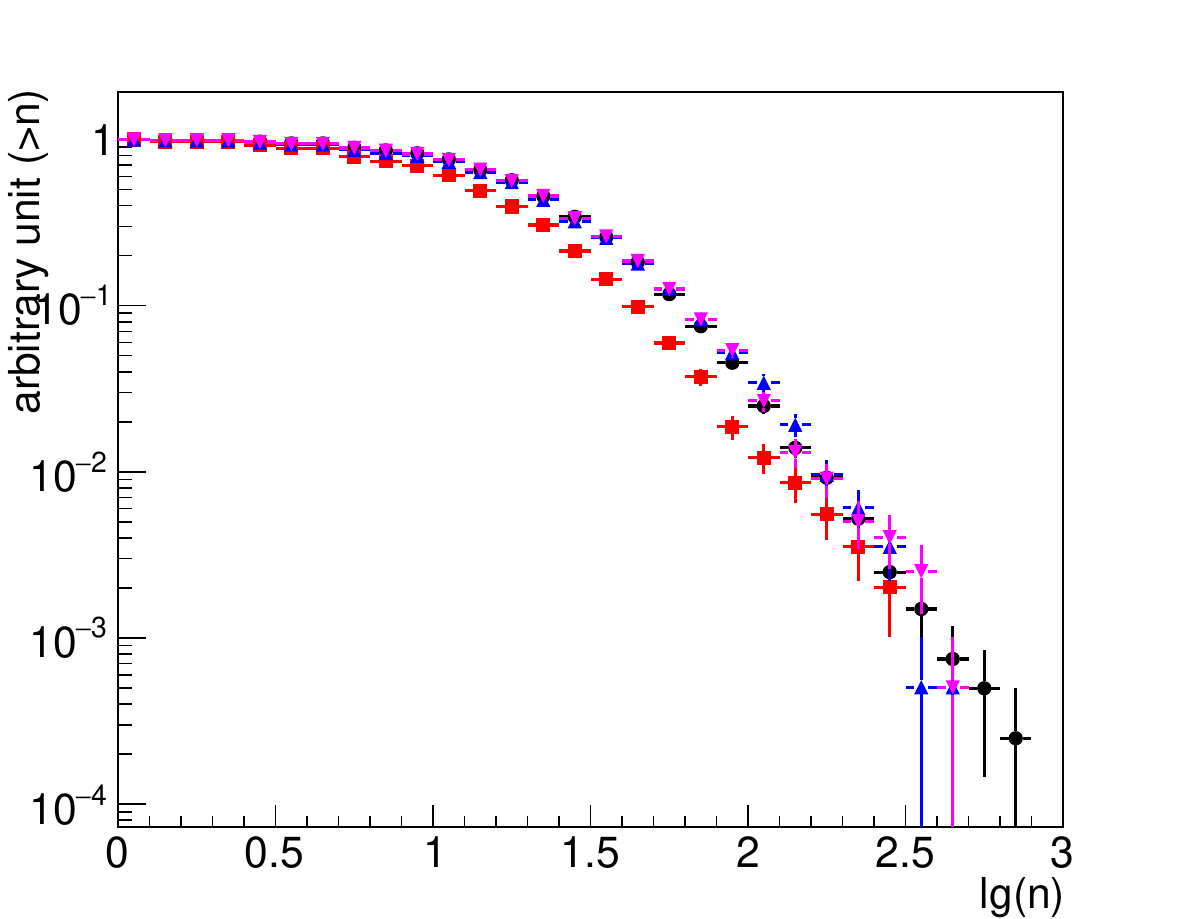}
	\\[1mm]
	\caption{
	The effect of different target materials on the thermal neutron integrated distributions normalized to the simulated numbers. Black dots are for case A, red squares for case B, purple anti-triangles for case C, and blue triangles for case D.
	}\label{fig:targets}

\end{figure}

Regarding optimization of the target material, the following four cases are simulated: \\
$\mathbf{A}$. all the EN-detectors are mounted on the ground, \\
$\mathbf{B}$. each EN-detector is mounted on a 1-m$^3$ sand cube (filled with sand) same as in the previous work~\cite{ENDA-ASS2022}, \\
$\mathbf{C}$. each EN-detector is mounted on a 4-m$^3$ (2 m $\times$ 2 m $\times$ 1 m) sand cube, and\\
$\mathbf{D}$. each EN-detector is mounted on a 1-m$^3$ graphite cube (filled with graphite). \\
Comparing the thermal neutron integrated distributions of the four cases (Fig~\ref{fig:targets}), we see that the case with the EN-detectors mounted on the ground (case A) collects more thermal neutrons than that on a 1-m$^3$ sand cube (case B), and it collects similar thermal neutrons as that on a 4-m$^3$ sand cube (case C). These results confirm that the 1-m$^3$ sand cubes cause a reduction in the thermal neutrons in the EAS events, mostly due to its geometrical factor being less than 1 and consequently the reduction in the target material. When the volume of the sand cubes increases to 4 m$^3$, the geometrical factor is increased to close to 1 so that the number of collected neutrons is close to that by the detectors mounted on the ground. Moreover, as the target material, sand and soil are similar in their thermal neutron collections, but because sand has more uniform components than soil, it is beneficial to obtain more precision detection result and keep consistency between the simulation and the real detection data. Moreover, in practice, sand cubes can also protect the target material from rain water which can reduce the thermal neutrons reaching the detectors. If the target material is replaced by 1-m$^3$ graphite (case D), the collected neutrons are still close to the cases when the detectors are mounted on the ground (case A) or on the 4-m$^3$ sand cubes (case C). This is because that carbon is much lighter than the atoms in sand or soil, such as O, Si, Ca, etc., so that more thermal neutrons are generated during elastic collisions with the evaporation neutrons. But graphite is much more expensive than sand (about 60 times). Also, the components of sand are much more uniform than soil. Based on the above considerations, we choose case C, i.e. with each EN-detector mounted on a 4-m$^3$ sand cube, as the most feasible option.

\subsection{The trigger efficiency}
\begin{figure*}[b!] 	
	\centering
	  \includegraphics[width=0.45\linewidth]{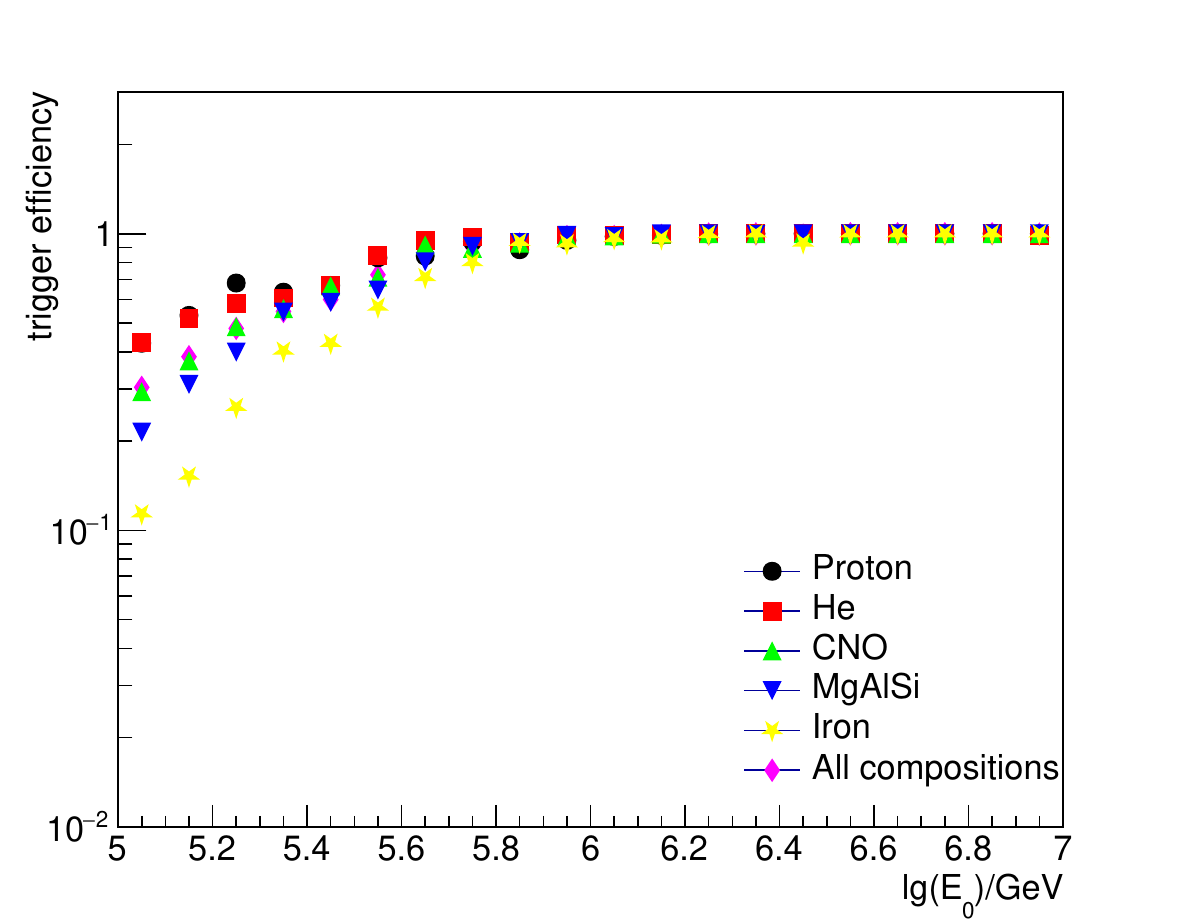}
   	  \includegraphics[width=0.45\linewidth]{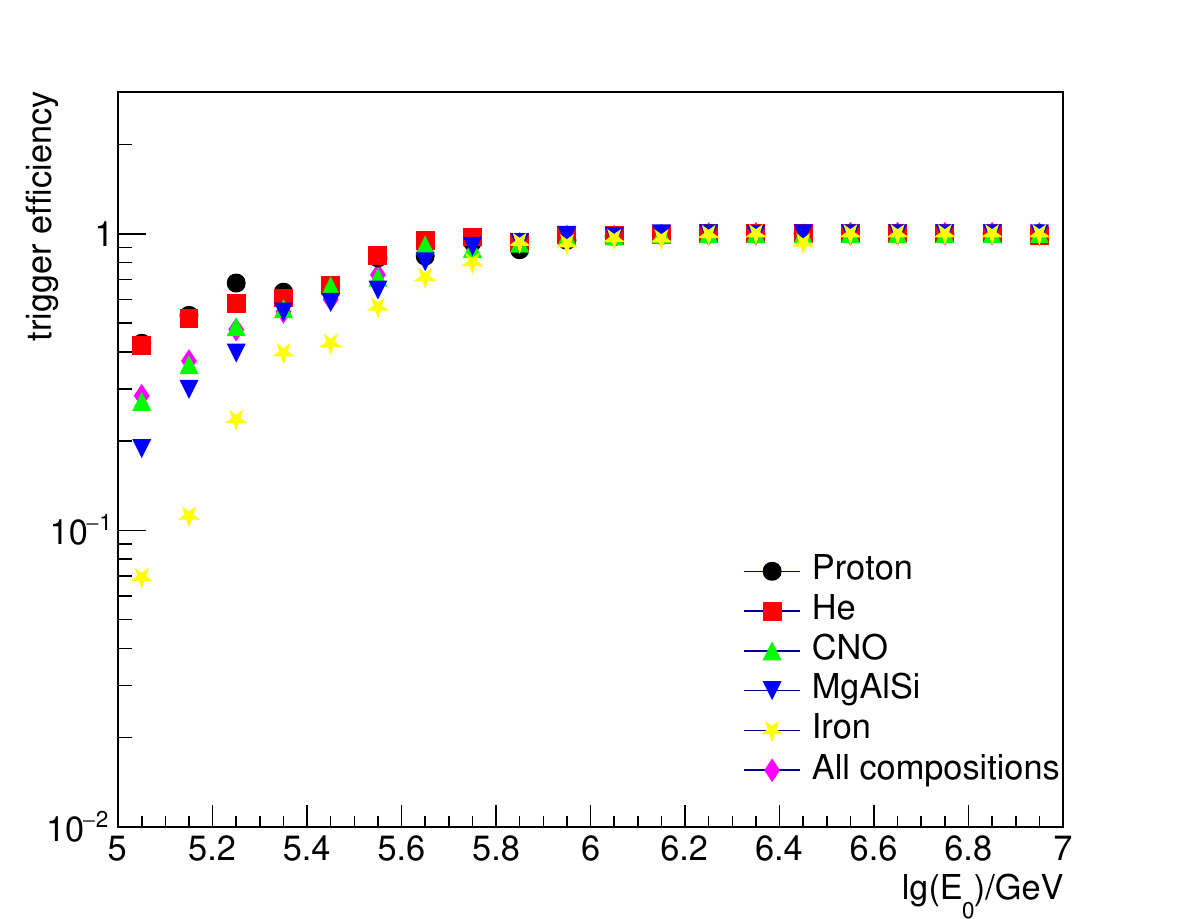}\hfil
    	\includegraphics[width=0.45\linewidth]{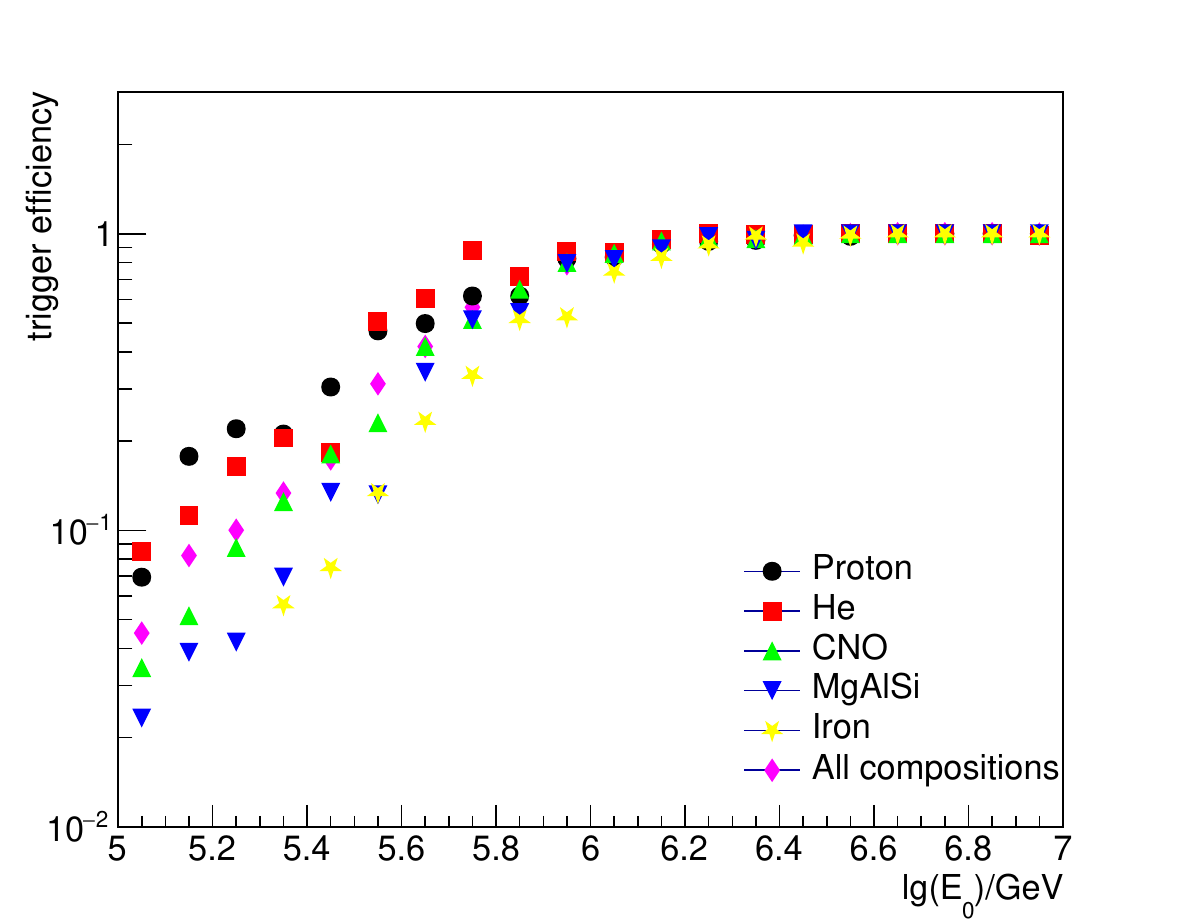}
    	\includegraphics[width=0.45\linewidth]{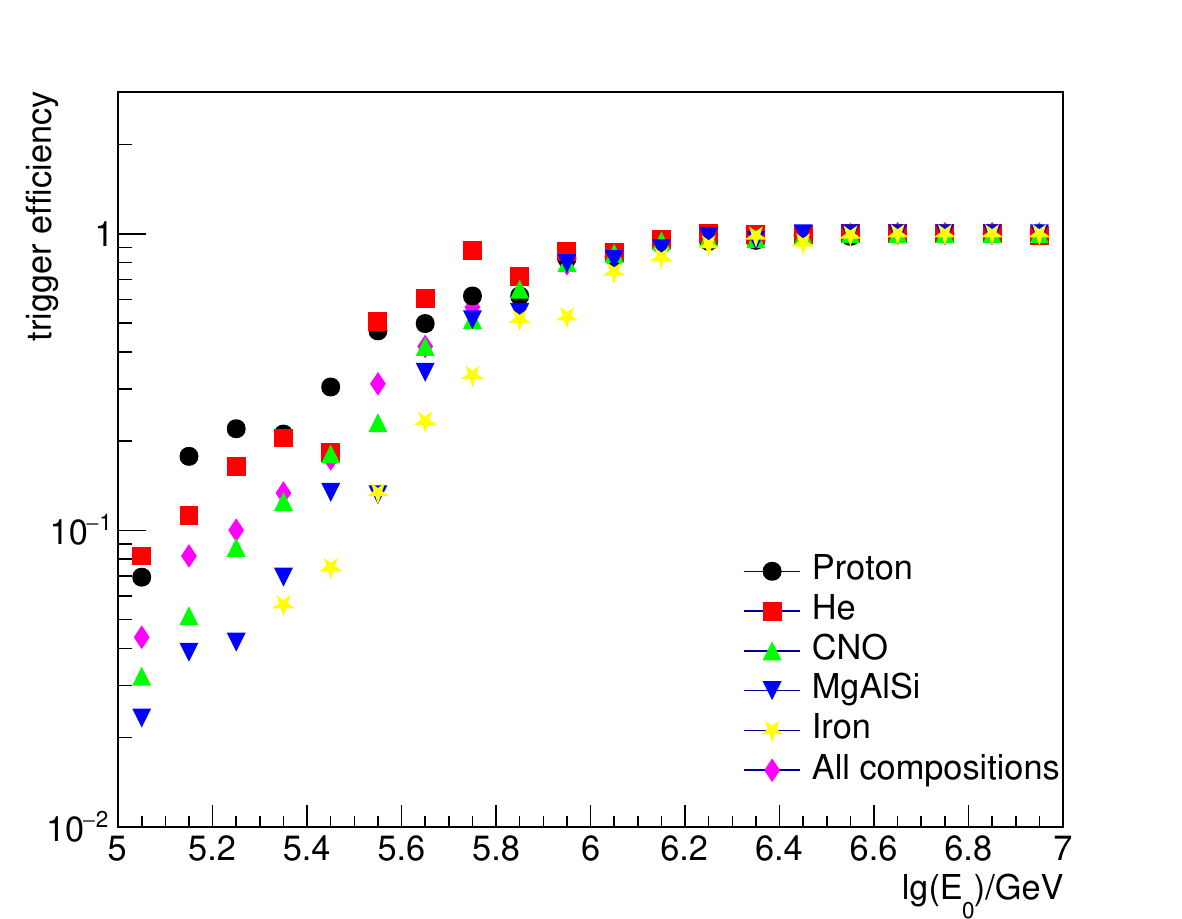}
	\\[1mm]
	\caption{
		The trigger efficiencies of the different trigger types at the energy range from 100 TeV to 10 PeV: (a) M1, (b) M2, (c) M3, and (d) M4. The markers represent different cosmic ray components: black circles for proton, red squares for He, light blue triangles for CNO, blue anti-triangles for MgAlSi, yellow stars for iron, and purple diamonds for all the components.
	}\label{fig:tig}

\end{figure*}

The trigger efficiencies by the different trigger types for different cosmic ray components at the energy range from 100 TeV to 10 PeV are obtained, as shown in (Fig.~\ref{fig:tig}). It can be seen that for the M1 and M2 trigger types (Fig. 6 (a) and (b)) which record lower energy electrons and neutrons, corresponding to the lower energy cosmic ray EAS, the trigger efficiency can reach above 90\% at the energy range of \textgreater 500 TeV. And for the M3 and M4 trigger types (Fig. 6 (c) and (d)) which record higher energy electrons and neutrons, corresponding to the higher energy cosmic ray EAS, the trigger efficiency can reach above 90\% at the energy range of \textgreater 1 PeV.

\subsection{The separation capability of cosmic ray composition }
The cosmic ray primary energy and compositions are determined by a combination of EAS measurements from both LHAASO and ENDA. The primary energy will be determined by using a parameter $N_{e\mu}$ combined by muons and electrons measured by LHAASO-KM2A~\cite{KM2A-E}. $N_{e\mu}$ can reduce the energy dependence on the cosmic ray components and the bias of individual compositions is made down to 5\%, and the obtained energy resolution is less than 10\% and the energy bias less than 5\% at the PeV range. Before analyzing LHAASO and ENDA coincident events, which is beyond the scope of this paper, we still use the true energy in a wide range, for example, from 1.5 PeV to 2 PeV, which is much larger than the energy resolution.

In this paper, in the fixed energy range, because of only using the ENDA detection, the cosmic ray primary compositions are separated by only using the thermal neutrons ($N_n$) and electrons (i.e., $m.i.p.$s in detection) which are the total thermal neutrons and m.i.p.s of the detector recording the most m.i.p.s and the 8 neighborhood detectors, respectively. The following event selection rules are used:

\begin{itemize}
    \item a shower is dropped in an area of 10 m$\times$10 m at the center of the detectors array,
    \item more than 6 detectors record \textgreater 4 m.i.p.s  for a shower, and
    \item in one single shower, the detector recording the most m.i.p.s is one of the 4 central detectors.
\end{itemize}

During the data analysis, it is found that, because of the lateral distributions of the thermal neutrons and electrons, both $N_n$ and $m.i.p.$s are anti-correlated with the distance between the true shower core and the detector observing the maximum $m.i.p.$s in the shower. As a consequence, in order to reduce the influence of the distance, the following parameters are taken for the component selection,

\begin{equation}
\Sigma_{n} = \Delta \times N_{n}
\label{Sigman}
\end{equation}
and
\begin{equation}
\Sigma_{e} = \Delta \times {m.i.p.s}
\label{Sigmae}
\end{equation}

where $\Delta$ is the distance between the shower core and the detector observing the maximum $m.i.p.$s in the shower. The shower core position is calculated in a weighted center method, by taking $m.i.p.$s as the weight, by which the core position resolution (containing 70\% of the events) reaches 1.3 m at the PeV range.

\begin{figure*}[ht!]
	\vspace*{1mm}
	\centering
	\includegraphics[width=0.34\linewidth]{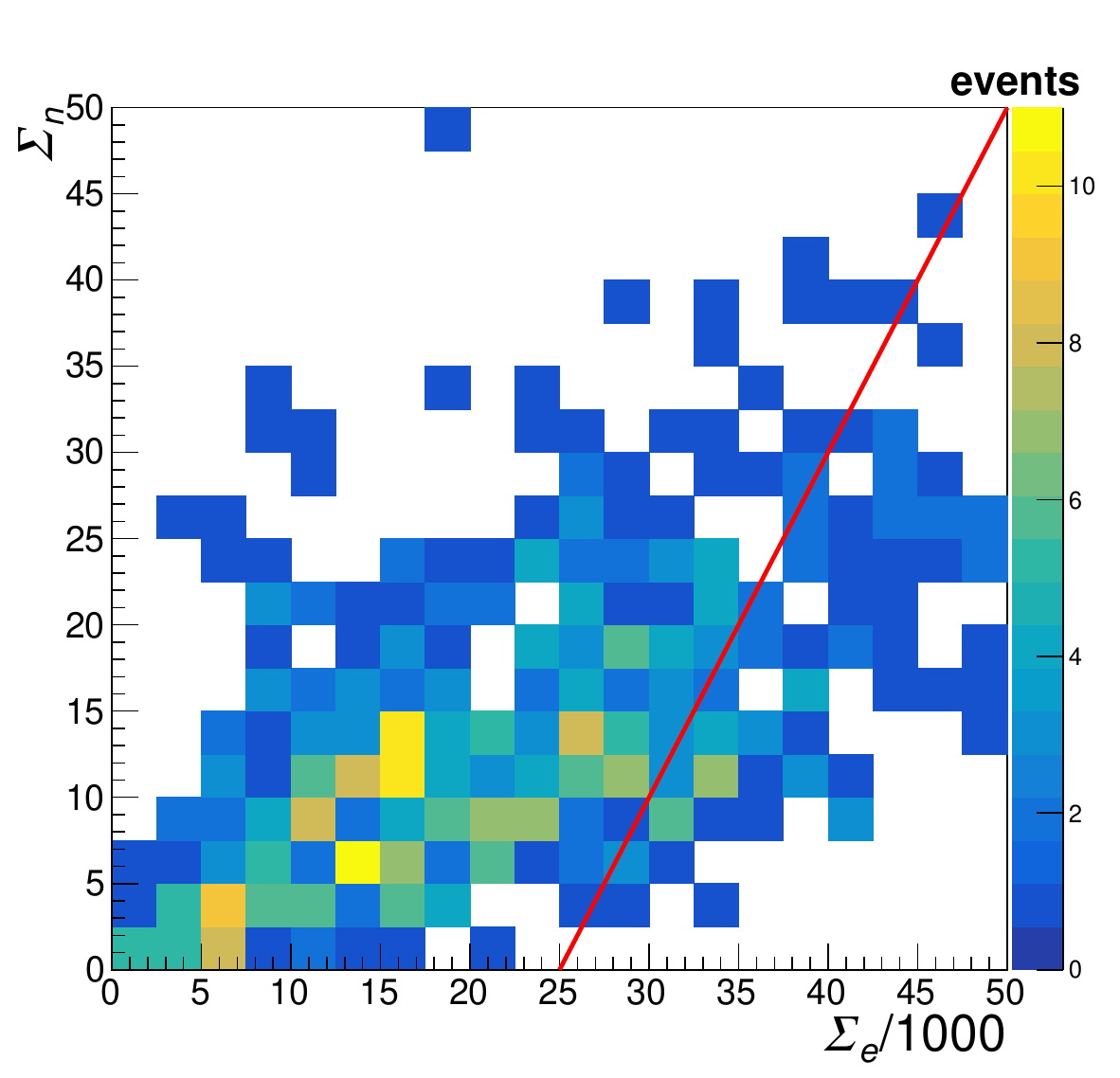}
	\includegraphics[width=0.34\linewidth]{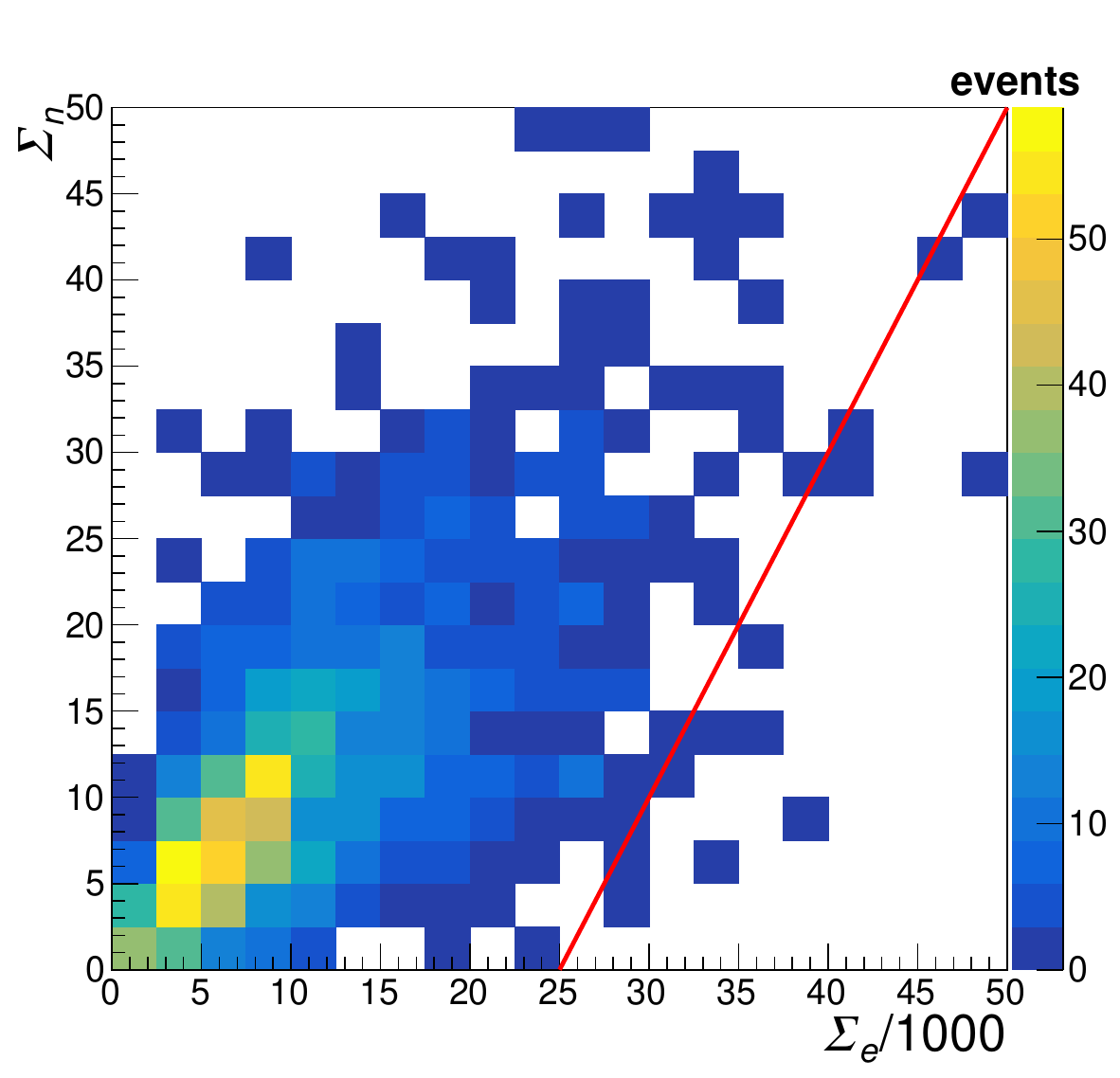}\hfil
	\includegraphics[width=0.34\linewidth]{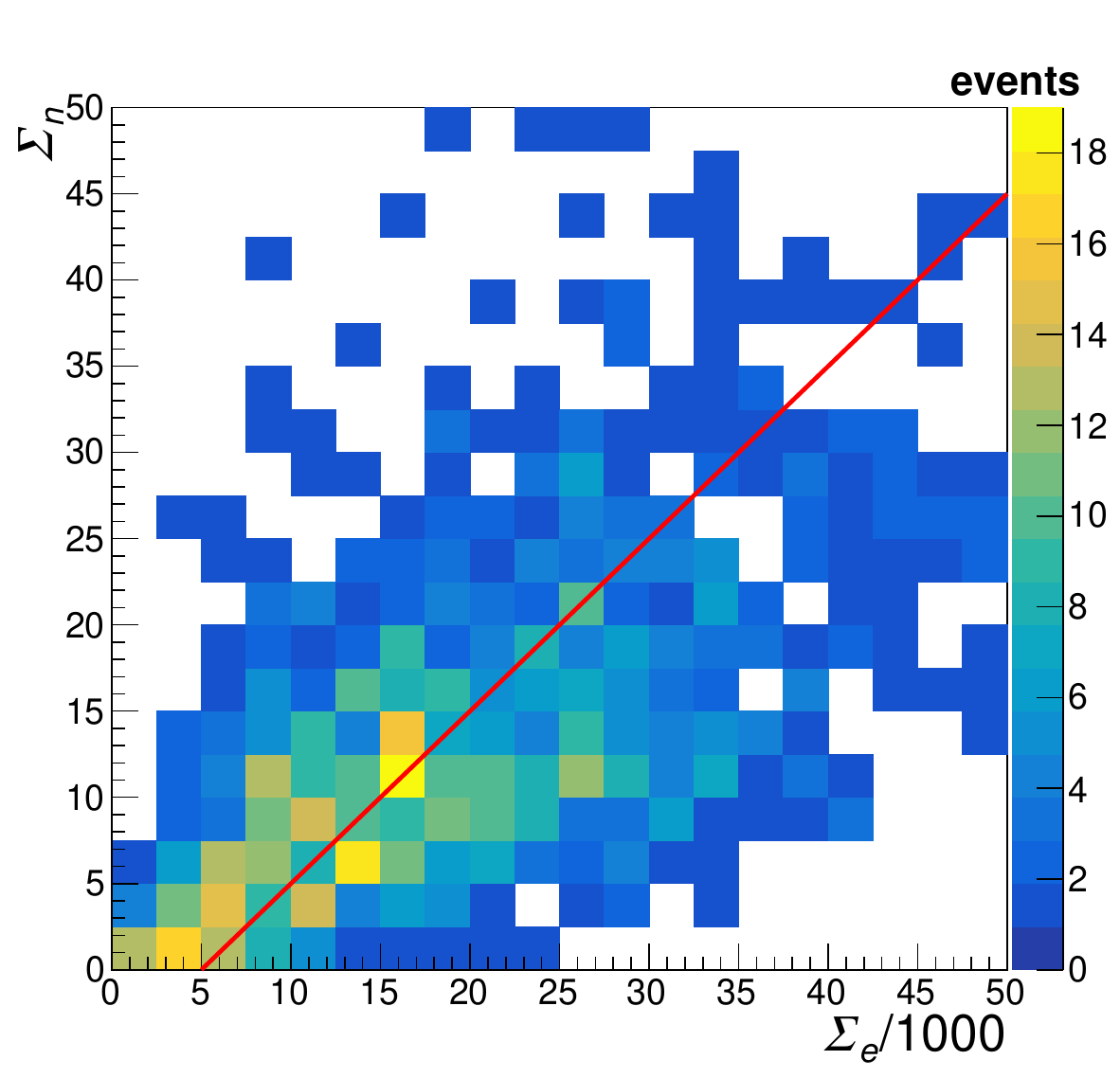}
	\includegraphics[width=0.34\linewidth]{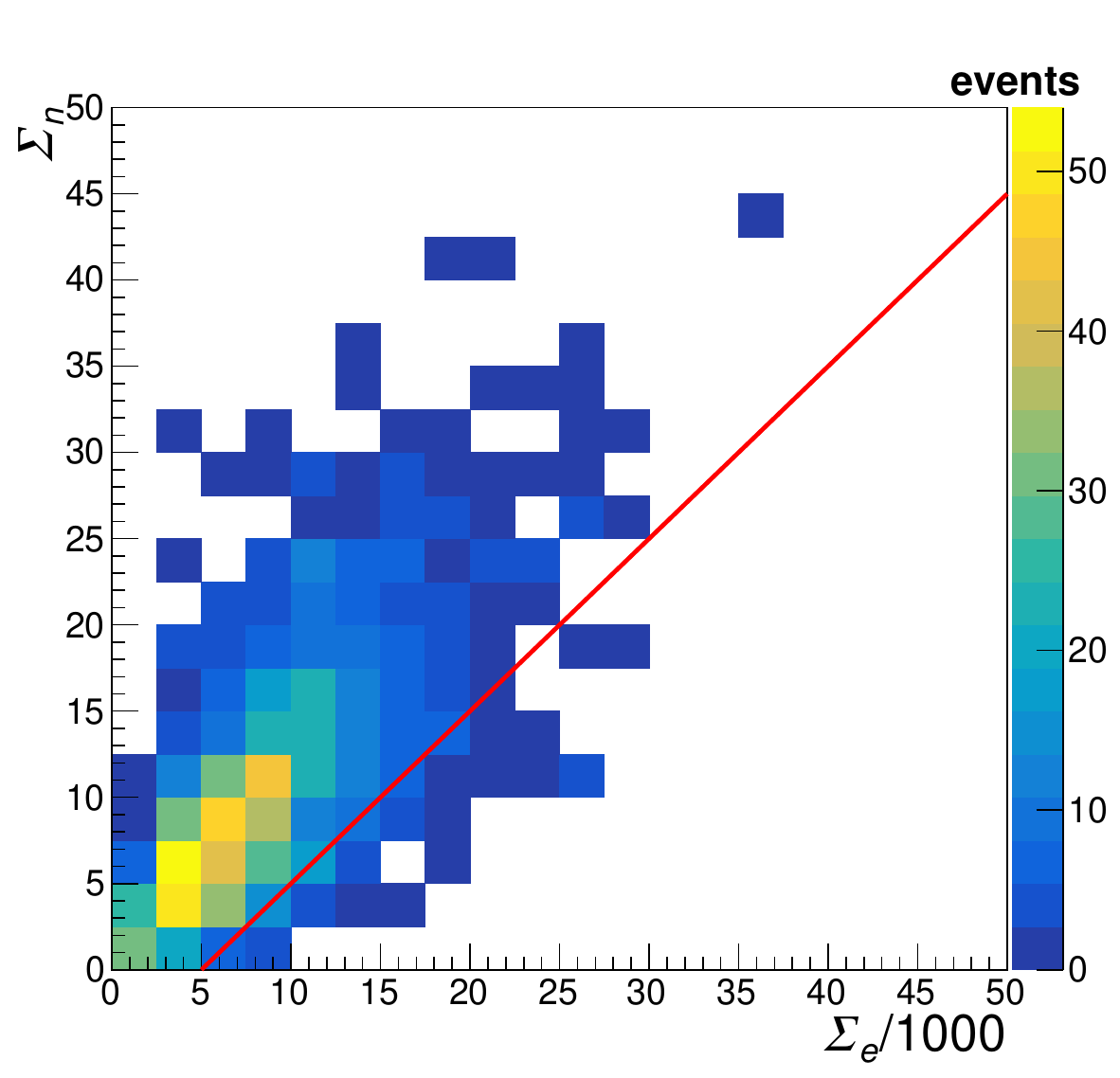}\hfil
	\includegraphics[width=0.34\linewidth]{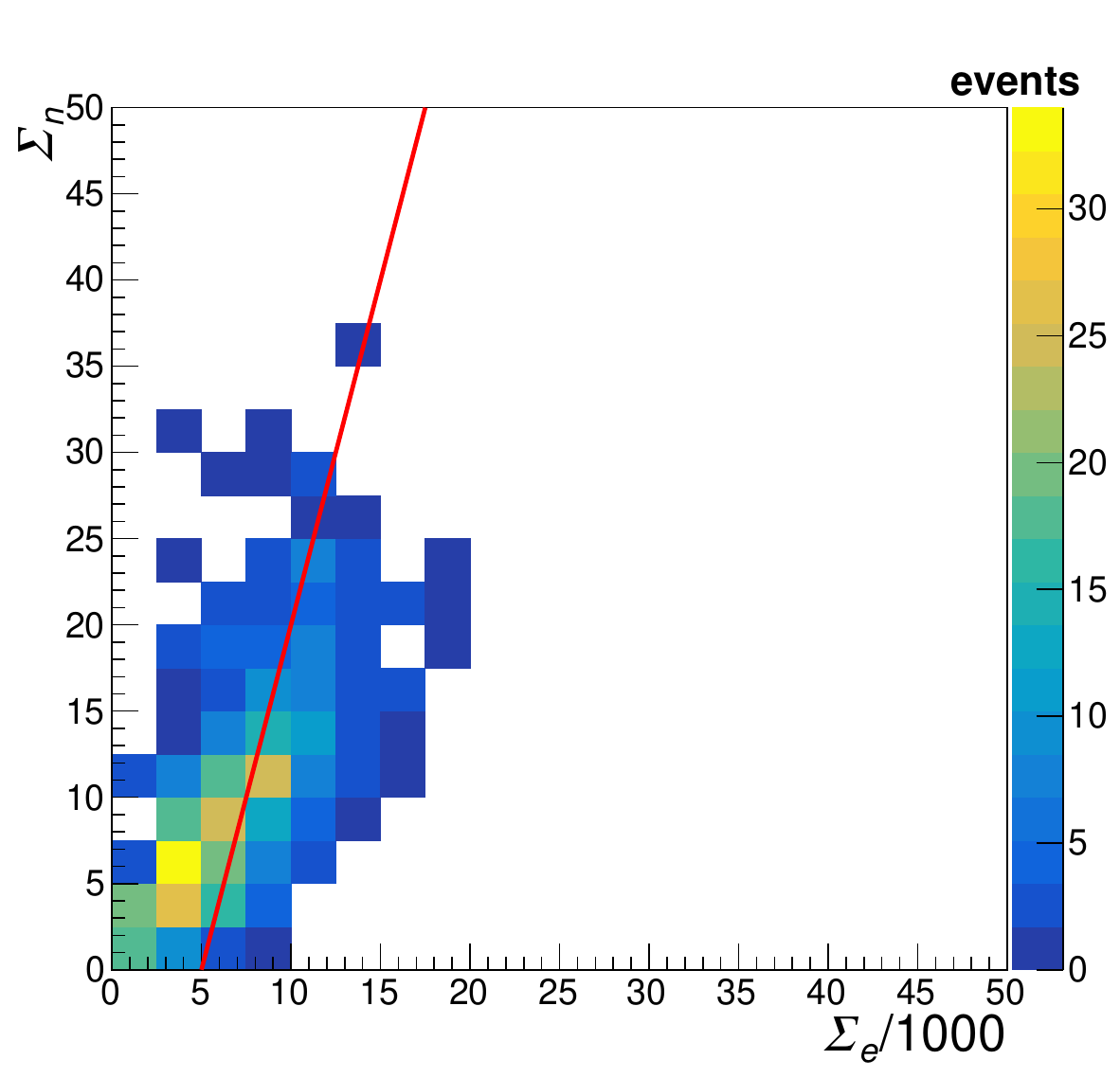}
	\includegraphics[width=0.34\linewidth]{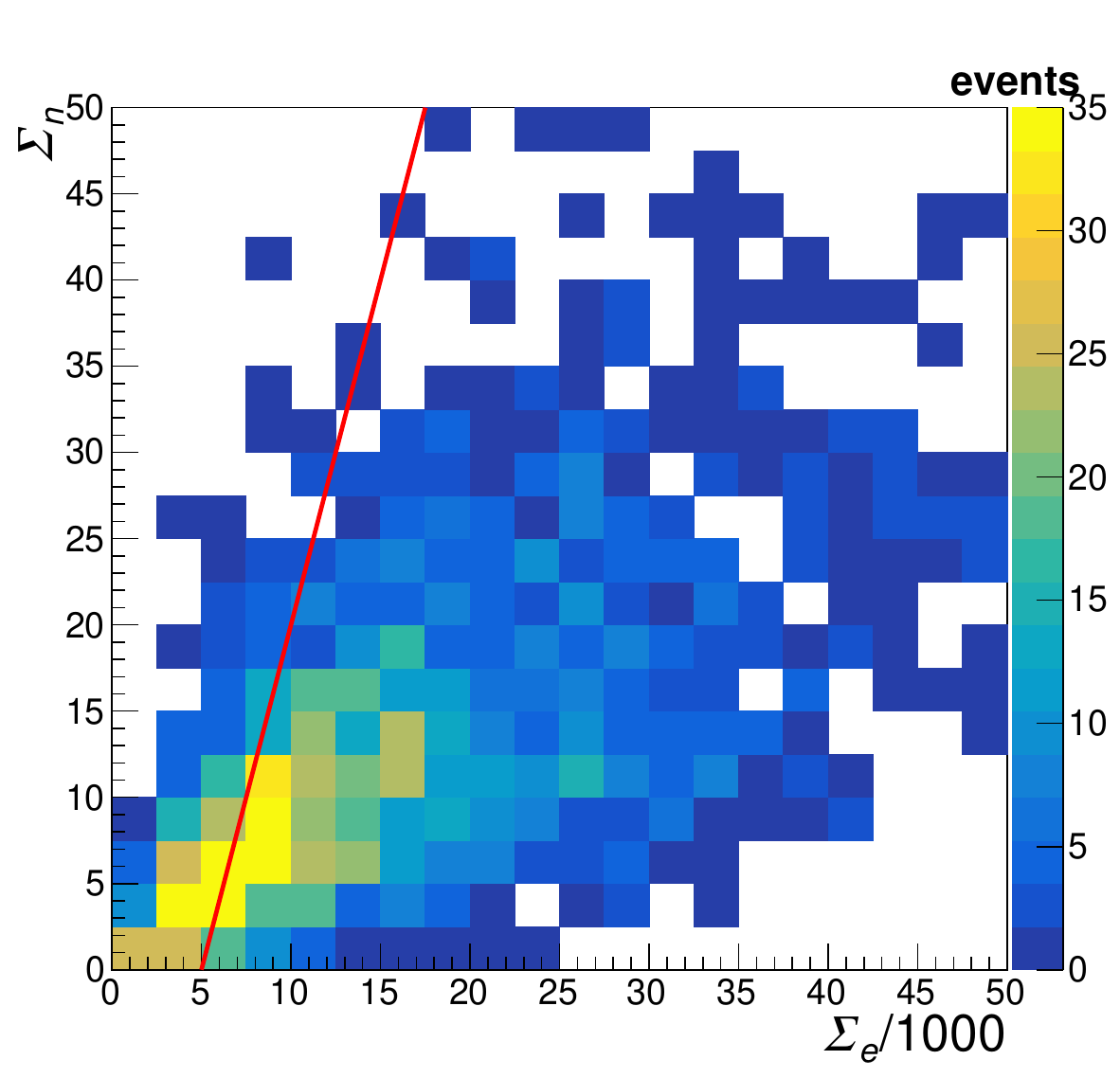}
	\\[1mm]
	\caption{The $\Sigma_e$ and $\Sigma_n$ distributions of different cosmic ray components in three component selections: (1) proton (a) and the others as contaminations (b), (2) light component (c) and the others as contaminations (d), (3) iron (e) and the others as contaminations (f). The red lines represent the component separation lines having the slope $a_0$ and the intersection $b_0$ obtained from Fig. \ref{a-b} and listed in Table \ref{tab2}.
	}\label{Nn-Ne}
	
\end{figure*}

\begin{figure*}[b!]  
	\vspace*{1mm}  
	\centering  
		\includegraphics[width=0.3\textwidth]{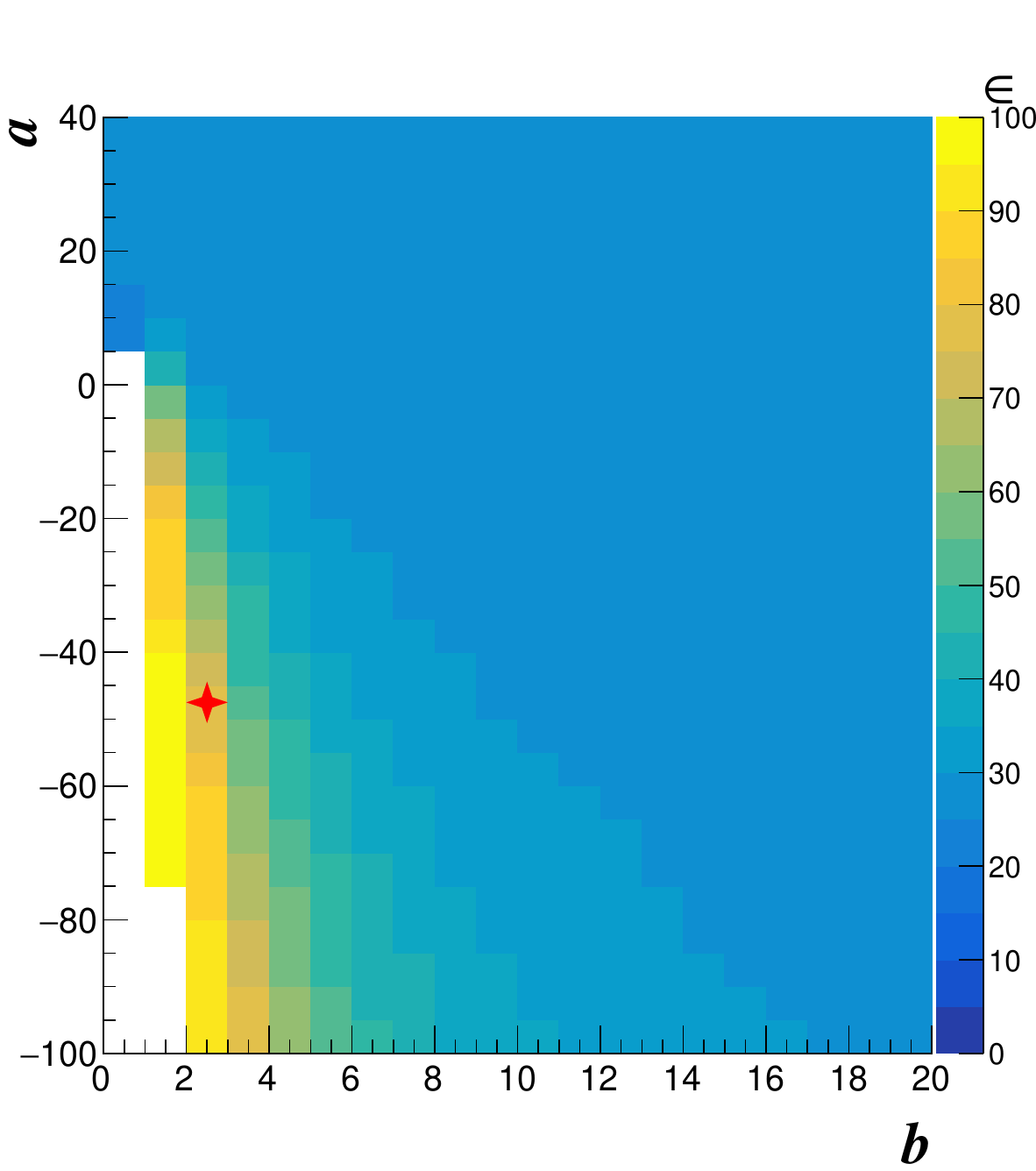}
		\includegraphics[width=0.3\textwidth]{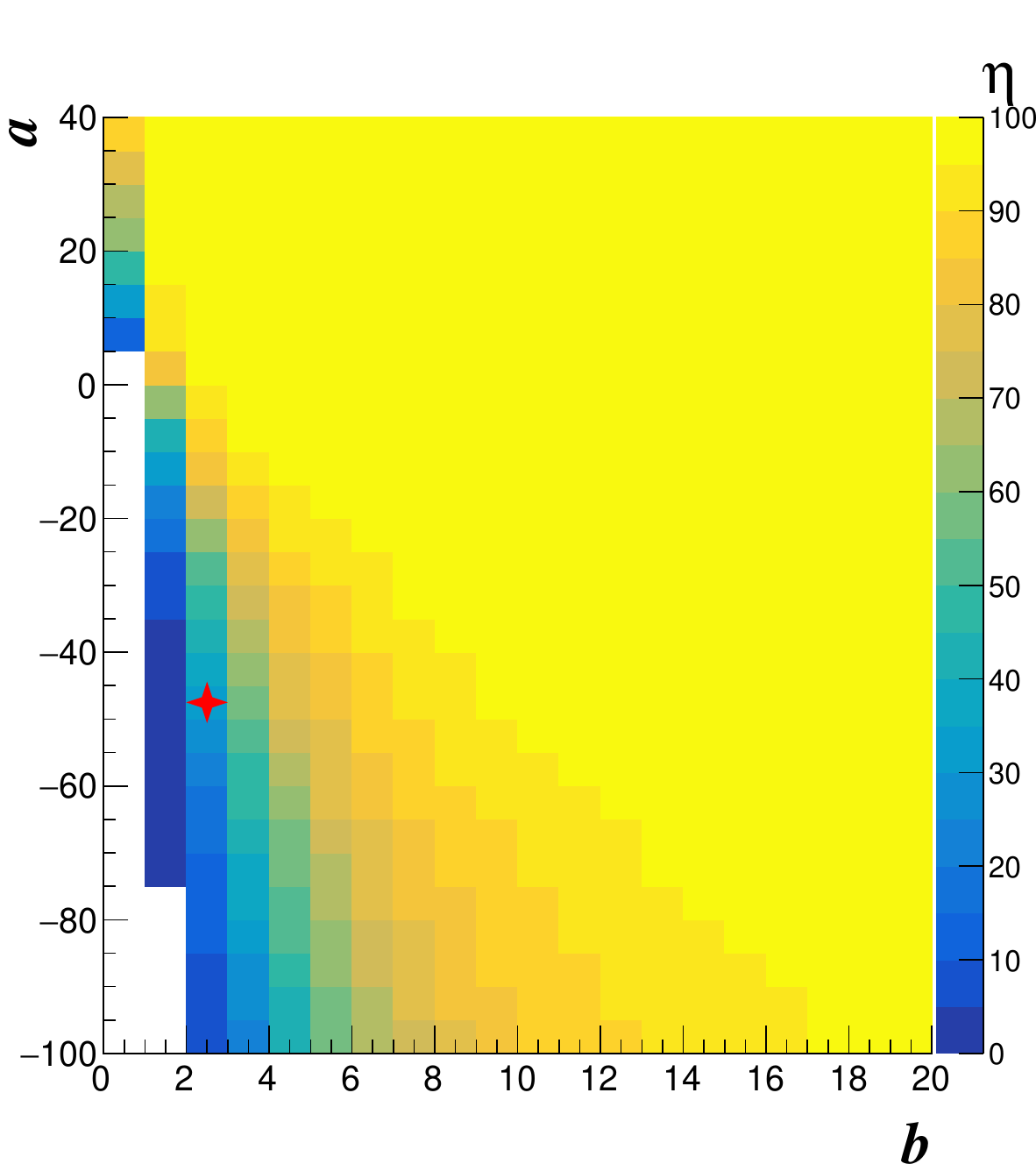}
		\includegraphics[width=0.3\textwidth]{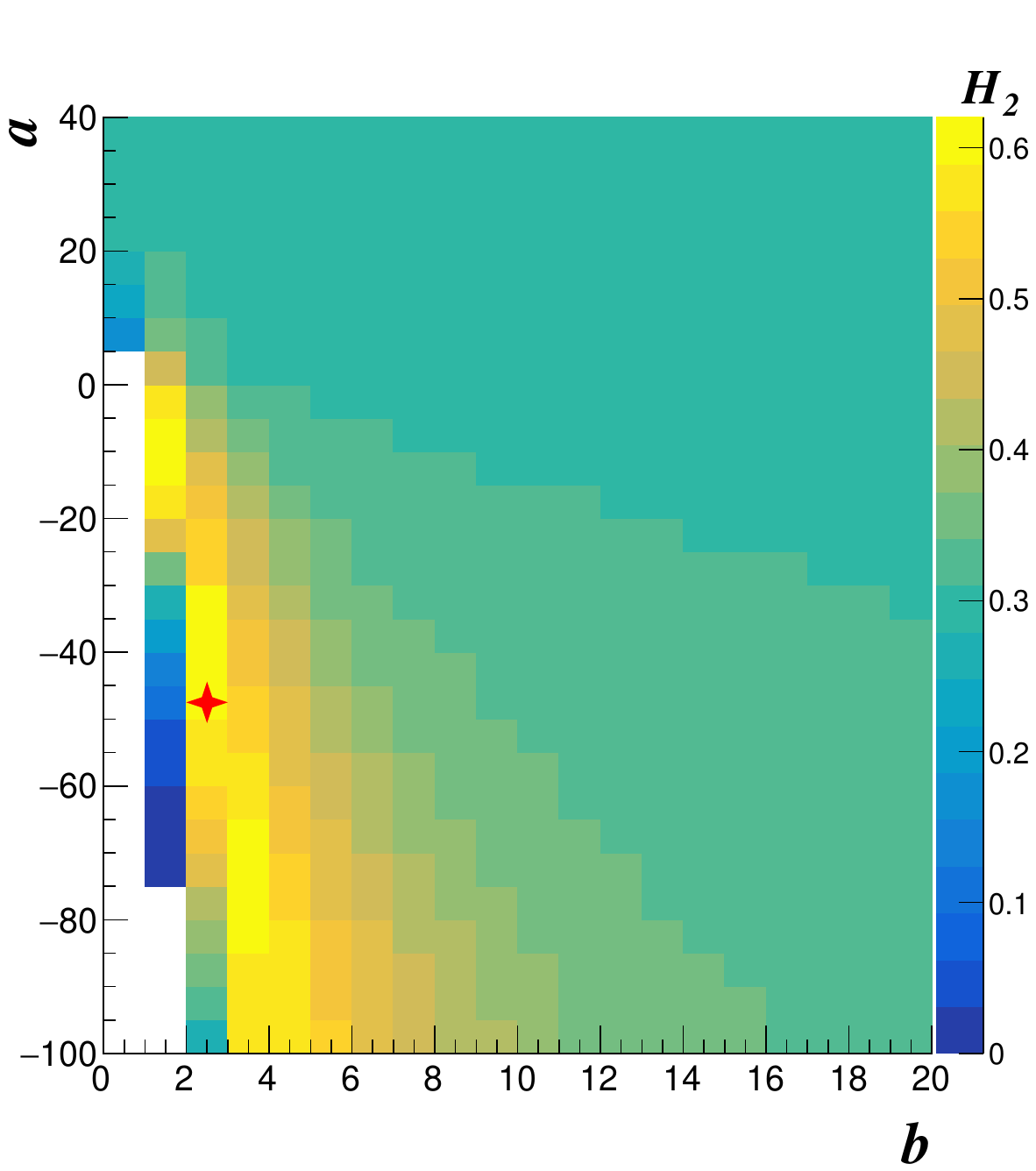}\hfil 
	\includegraphics[width=0.3\textwidth]{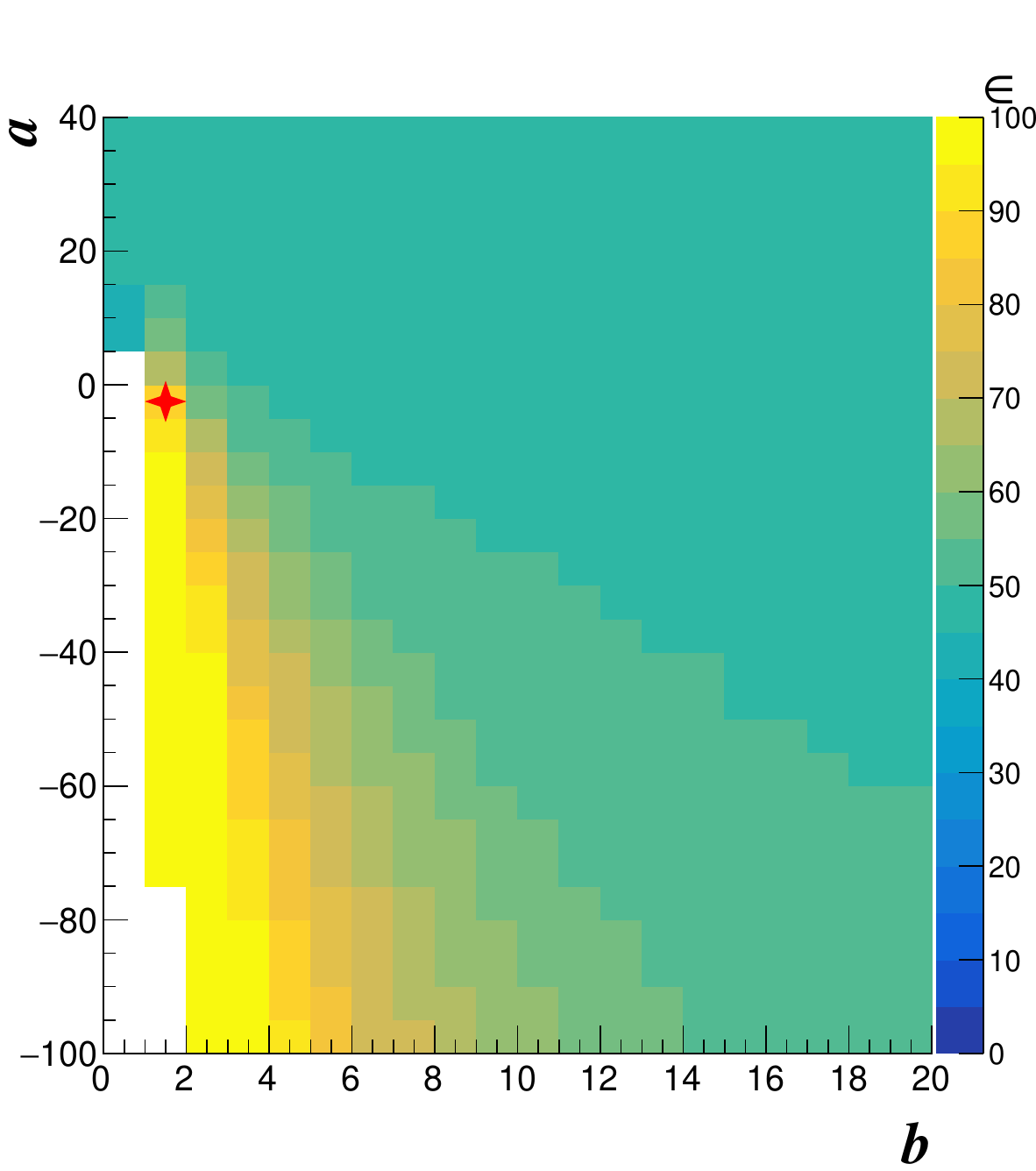}
	\includegraphics[width=0.3\textwidth]{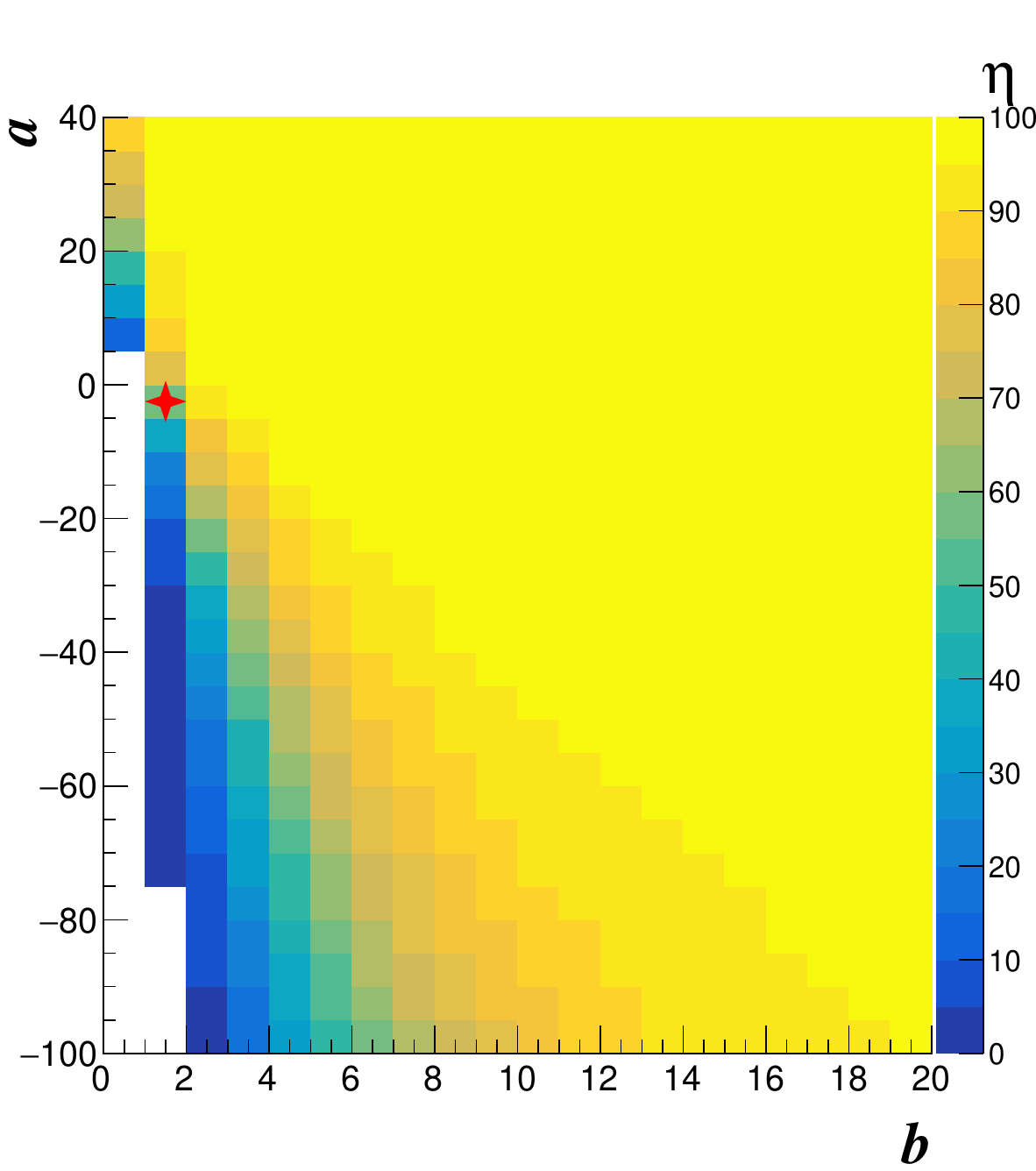}
	\includegraphics[width=0.3\textwidth]{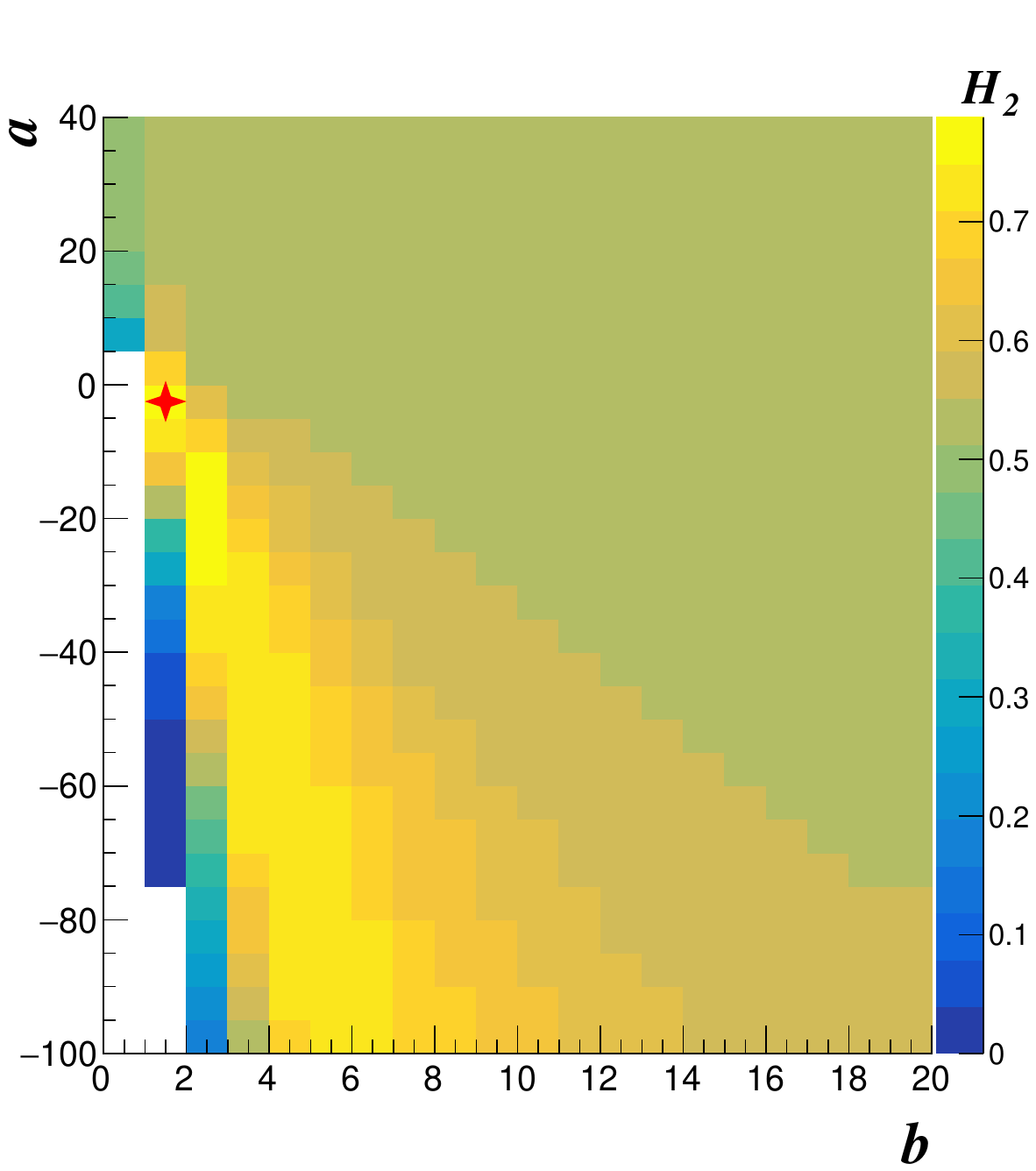}\hfil
	\includegraphics[width=0.3\textwidth]{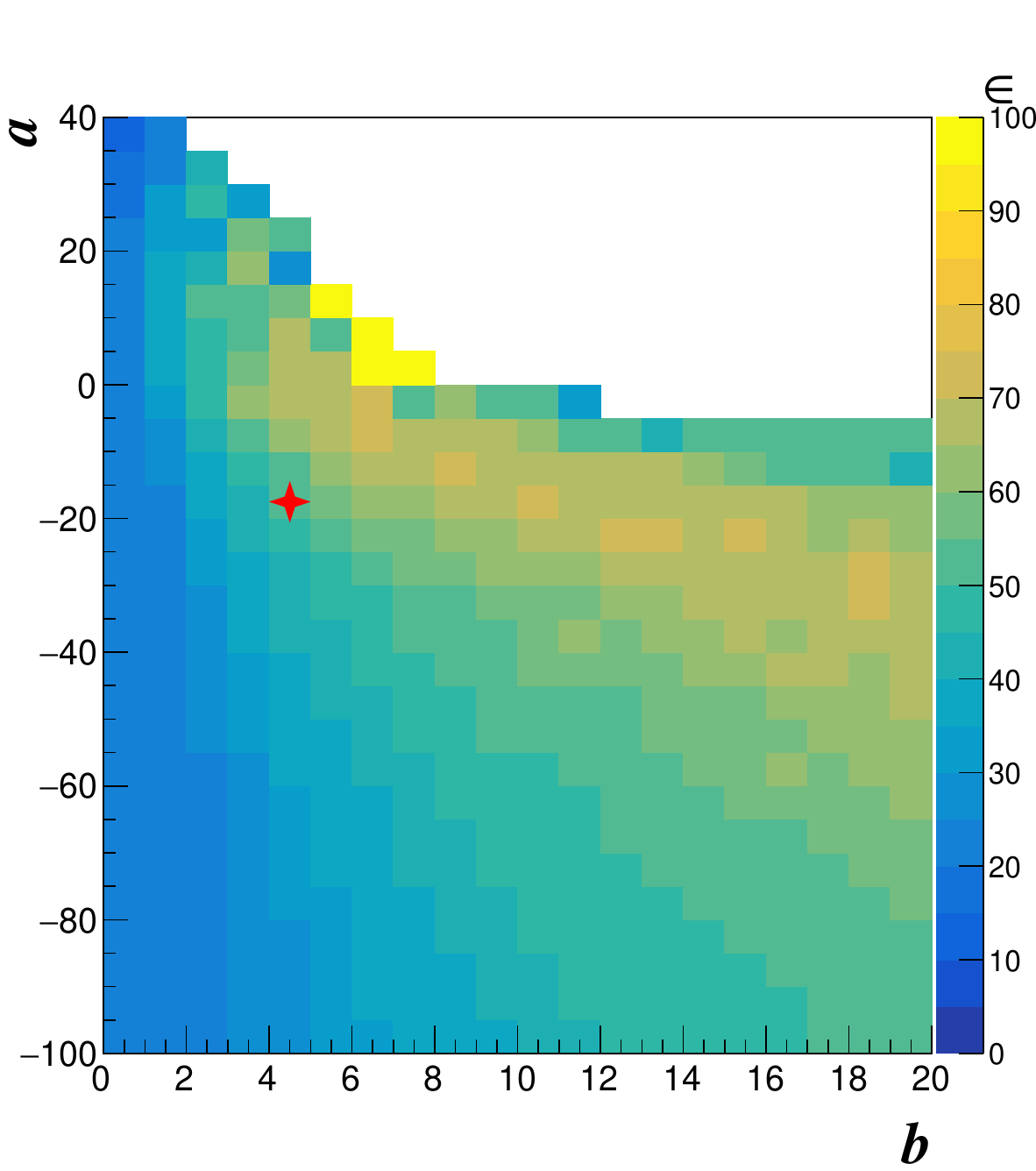}
	\includegraphics[width=0.3\textwidth]{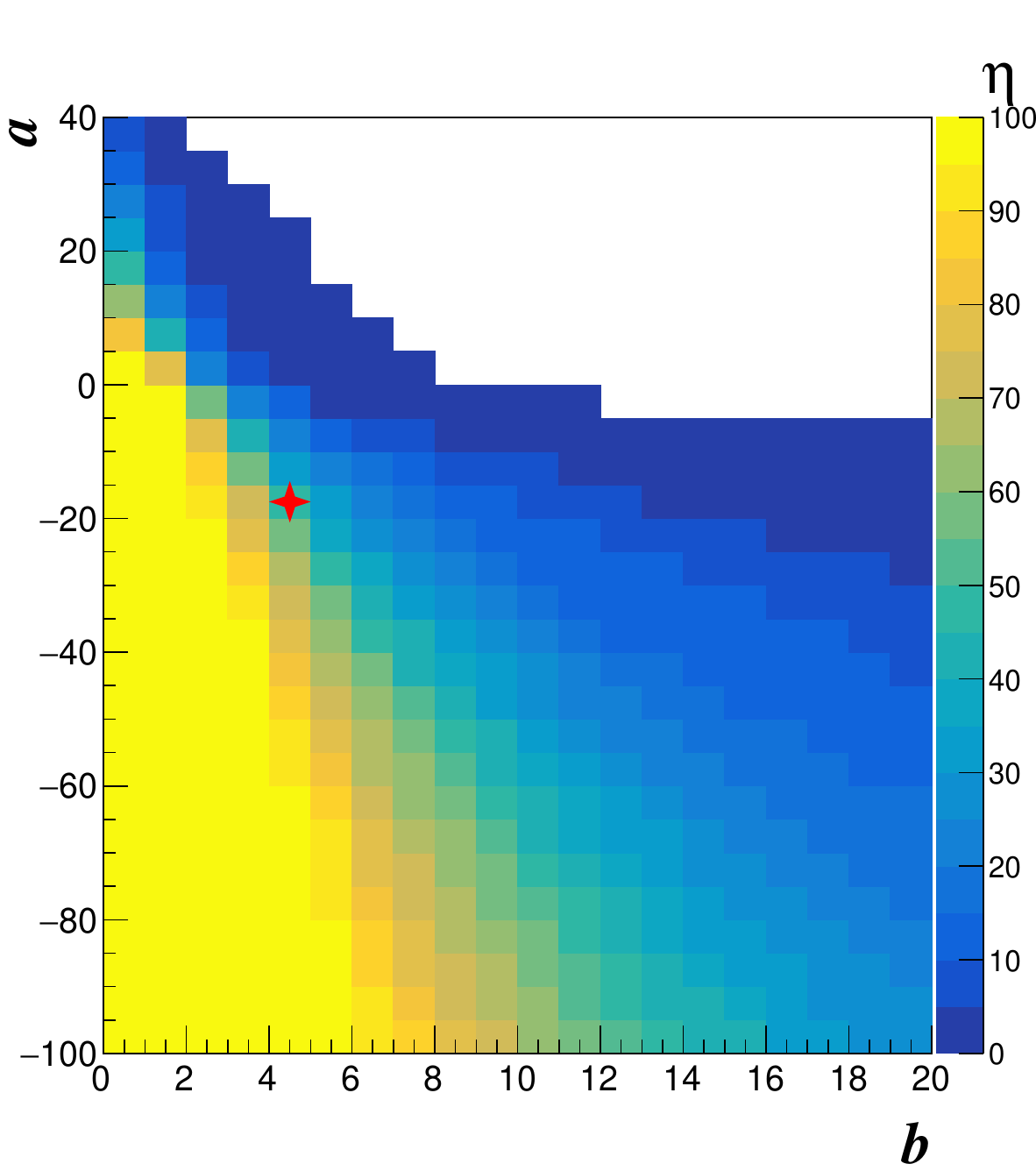}
	\includegraphics[width=0.3\textwidth]{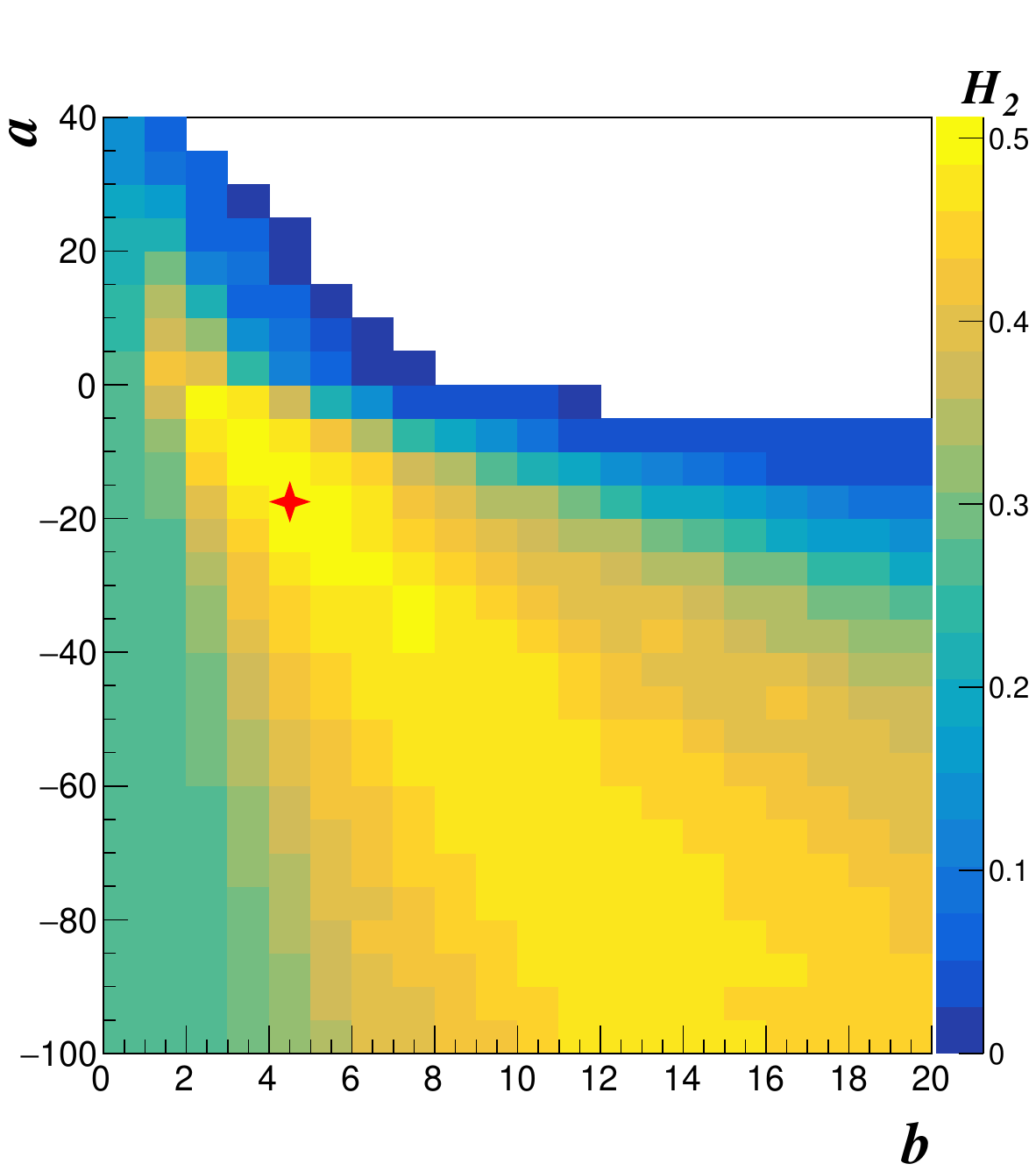}
	\\[1mm]
	\caption{The values of $\epsilon$, $\eta$, and $H_2$ at various $b$ and $a$ values for the three target components: (1) $\epsilon$ (a), $\eta$ (b), and $H_2$ (c) for the target proton, (2) $\epsilon$ (d), $\eta$ (e), and $H_2$ (f) for the target light component, and (3) $\epsilon$ (g), $\eta$ (h), and $H_2$ (i) for the target iron. The white place in the $\epsilon$ and $\eta$ plots is where there is no any event selected, and that in the $H_2$ plots is where both $\epsilon$ and $\eta$ are 0. The red crosses are the optimal values of $a_0$ and $b_0$ for the target components.
	}\label{a-b} ,
	\vspace*{-3mm}
\end{figure*}

During the component selections, three kinds of target components are considered: (1) proton, (2) light component (proton + Helium), and (3) iron. For different components, the distributions of the thermal neutrons and $m.i.p.$s are shown in Fig.~\ref{Nn-Ne}, where three pairs of plots are for the three kinds of target components, respectively: in (a) and (b), the events are selected as proton on the right of the component separation line; in (c) and (d), the events are selected as light component on the right of the component separation line; and in (e) and (f), the events are selected as iron on the left of the component separation line. One component separation line is a straight line having two parameters: the intersection $a$ and the slope $b$. When $a$ and $b$ are certain, the following three parameters are determined,
\begin{itemize}
    \item the purity ($\epsilon$) which is the ratio of the events of the selected target component to the events of contamination,
    \item the selection efficiency ($\eta$) which is the ratio of the events of the selected target component to the total events of the target component, and
    \item the weighted harmonic mean ($H_2$) is defined as,
    \begin{equation}
     H_2 = \frac {(w_{\epsilon}+w_{\eta})\eta\epsilon}{w_{\eta}\epsilon+w_{\epsilon}\eta}
    \label{qf}
    \end{equation}
    where $w_{\epsilon}$ is the weight of purity ($\epsilon$) and $w_{\eta}$ is the weight of selection efficiency ($\eta$). In order to obtain $\epsilon$ with priority, we set $w_{\epsilon}=5$ and $w_{\eta}=1$.
\end{itemize}
The values of $\epsilon$, $\eta$, and $H_2$ at various b and a for three kinds of target components are drawn in Fig. \ref{a-b}. $H_2$ can be used as the judgement on the optimal $a$ and $b$ with considering compromise between $\epsilon$ and $\eta$. Finally, at the maximum value of $H_2$, the optimal values $a_0$ and $b_0$ for the three target components are selected and listed in Table \ref{tab2} and drawn as red crosses in Fig. \ref{a-b}, and the component  separation lines having $a_0$ and $b_0$ are also drawn in Fig. \ref{Nn-Ne}.

\begin{table}[b!]
	\centering
	\caption{The optimal values $a_0$ and $b_0$ and the corresponding $\epsilon$, $\eta$, and $H_2$ in the three component selections.}
	\begin{tabular}{ cc c c c c c}
		\midrule
		Target component   & $a_0$ & $b_0$ & $\epsilon$ & $\eta$ & $H_2$\\
		\hline
		Proton        & -50 & 2.0 & 76\% & 32\% & 0.62\\
	     Light component & -5.0 & 1.0 & 86\% & 56\% & 0.79\\
		Iron          & -20 & 4.0 & 52\% & 47\% & 0.51\\
		\midrule
	\end{tabular}
	\label{tab2}
\end{table}

\subsection{Comparison between experiment and simulation}
\begin{figure}[ht!]
 	\vspace*{1mm}
 	\centering
 		\includegraphics[width=9cm,height=7.2cm]{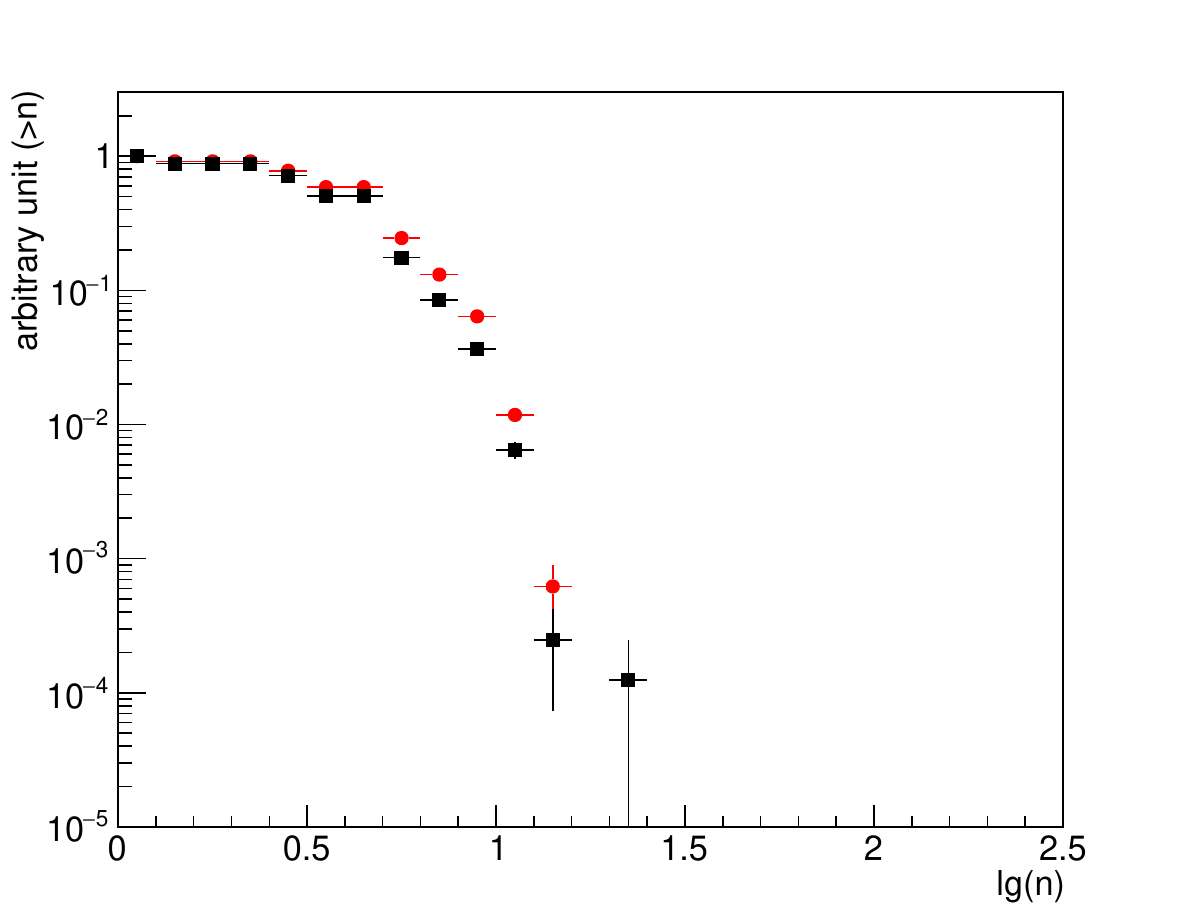}
 	\\[1mm]
 	\caption{The thermal neutron background integrated distributions normalized to the number of events from the trigger type M0 in one month in the PRISMA-YBJ-16 experiment~\cite{ENDA-ASS2022}. Red dots are for the detectors mounted on the ground, and black squares for the detectors mounted on the sand cubes.
 	}\label{fig:M0}
 \end{figure}

In the previous work~\cite{ENDA-ASS2022}, the so called ``sand cubes" each with a dimension of 1-m$^3$ were used to mount the detectors on top of them, and the data obtained by placing the detectors on or away from the sand cubes were compared to study the influence of the target material, which is a major environmental factor affecting the performance of the detector array. The composition of sand samples filled in the sand cubes and the soil under the array are tested carefully, and taken as the input for the corresponding material parameters in the simulation. Although the background neutrons can not be simulated, they can be taken from the trigger type M0 real data measured in the real experiments~\cite{ENDA-ASS2022}. The neutron distribution of M0 events in the PRISMA-YBJ-16 for one month is shown in Fig~\ref{fig:M0}. For each simulated EAS event, the total neutrons are the sum of the thermal neutrons generated from the real shower and the background neutrons which are randomly sampled from the neutron distribution of M0 events. Besides, for the total simulated EAS events, the background events with the same neutron distribution as the M0 events are added with normalization to the equivalent period of the real data.

\begin{figure}[ht!]
	\vspace*{1mm}
	\centering
		\includegraphics[width=9cm,height=7.2cm]{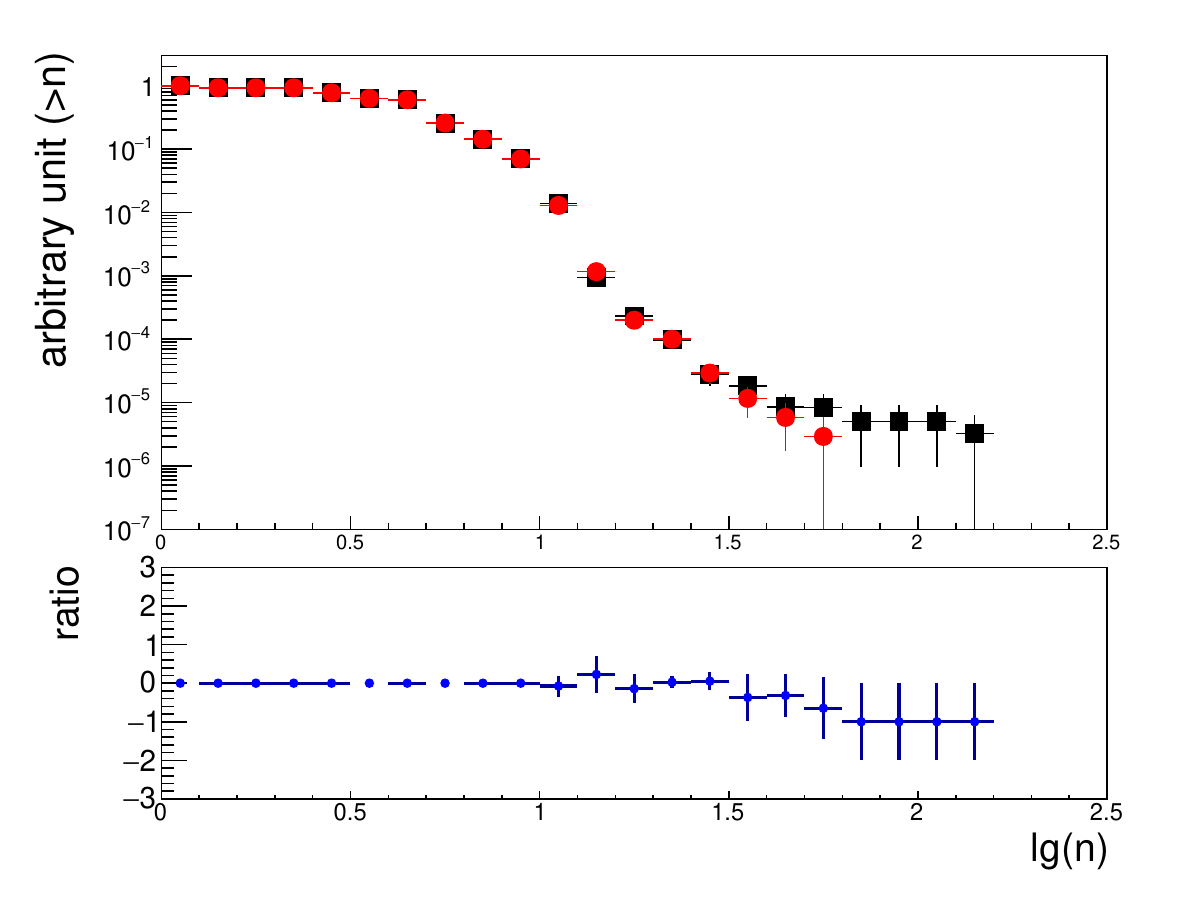}
		\includegraphics[width=9cm,height=7.2cm]{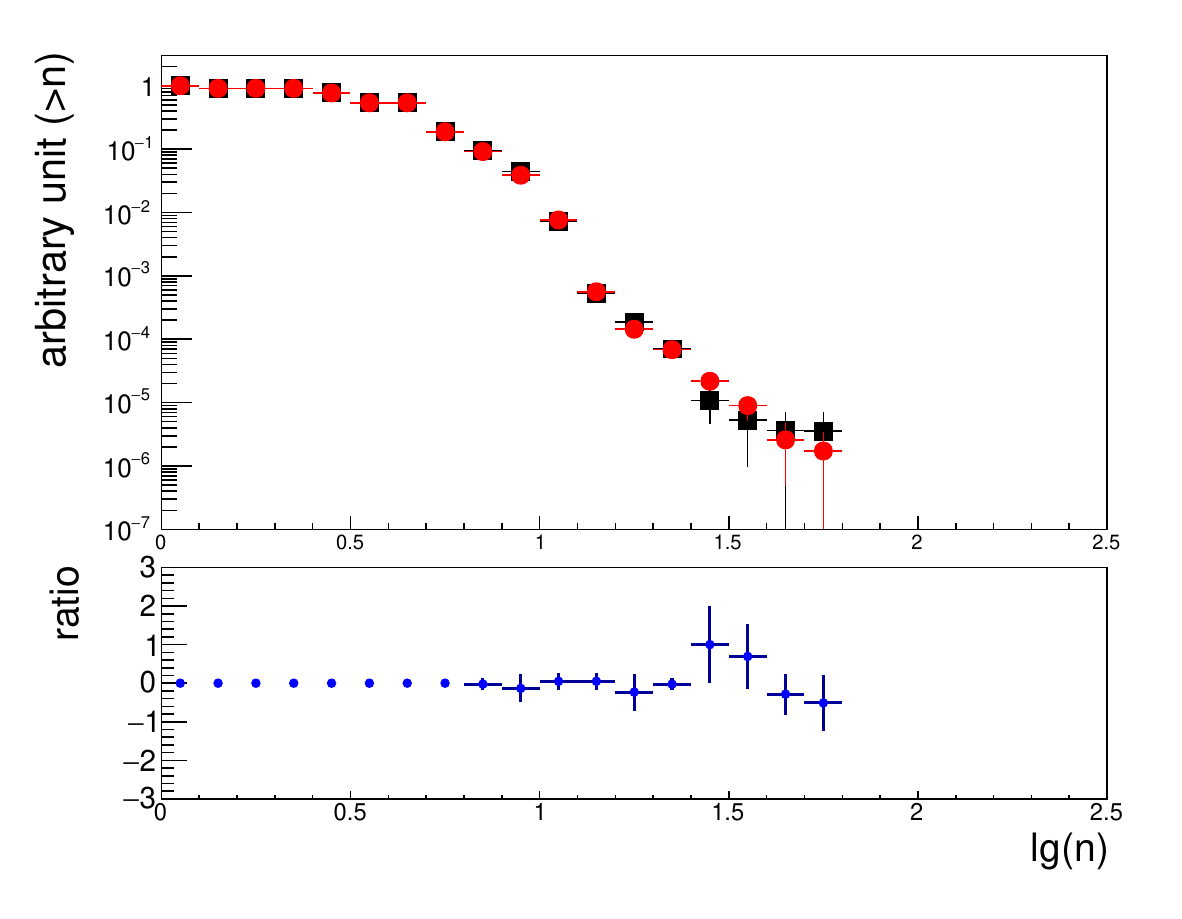}\hfil
	\\[1mm]
	\caption{
		  The thermal neutron integrated distributions normalized to the number of events for the Monte Carol simulation (red dots) and the experimental data (black squares)  with ratios of difference between the simulation data and the experimental data to the experimental data. Upper panel - without the sand cubes, and lower panel - with the sand cubes.
	}\label{fig:compare}
\end{figure}

Thermal neutron integrated distributions normalized to number of events of the simulation with sand cubes and without sand cubes are obtained and compared with ones of the real data respectively (Fig.~\ref{fig:compare}). In comparison, it is indicated that distributions in simulation in both case with sand cubes and case without sand cubes are consistent with the real data at neutrons less than 50. At neutrons larger than 50, the real data with sand cubes have three events more than the simulation due to fluctuation caused by low statistics at the highest energy. Consistency in the results between the simulation and the real data confirms the previous conclusions: even though the target materials have somewhat different compositions, changing target material by using sand cubes has only a small effect on detection efficiency. Moreover, for the array, due to reduction of target material, the sand cubes cause a reduction in thermal neutrons in EAS events, mostly due to geometrical factor while affect of soil chemical composition is not significant.

\section{Conclusion}
The ``knee region" problem in the cosmic ray energy spectrum is a hot and challenging topic in the cosmic ray research. It links to the centuries-old question on the origin, acceleration, and propagation of cosmic rays. Accurate measurement of hadrons in the shower core region is the key to solve the ``knee" problem, and is a probe to study the origin of cosmic rays. The EN-detector is a novel type of scintillator detector that can measure the core hadrons through measuring the thermal neutrons, which can be used to estimate the original composition and energy spectrum of cosmic rays. In LHAASO, KM2A measures the lateral distributions of electrons and muons, WCDA measures muons at far distance from the shower core, WFCTA measures atmospheric Cherenkov light to study the longitudinal development characteristics of the shower, and ENDA detects both thermal neutrons to obtain the information about the hadrons and the lateral distribution of electrons near the shower core.

It indicates consistency in the thermal neutron distributions between the real data obtained by the EN-detectors in the experiment and that in the Monte Carlo simulation. By comparing the results obtained at different adjacent detector distances and target materials, the detector array configuration is optimized. For the events recording higher electrons and thermal neutrons, which corresponds to higher energy cosmic ray EAS, the trigger efficiency can reach above 90\% at the energy range of >1 PeV. The capability of the EN-detectors in the cosmic ray composition separation by measuring the thermal neutrons generated is verified. In future studies, the current ENDA-64 is planned to be extended to the ENDA-400 composed by 400 EN-detectors to measure the  knee region of heavy components (Fe). Immediate follow-up research will focus on the simulation of the hybrid detection on both the composition and the energy spectrum of cosmic rays by combining the full secondary particle measurement of the EAS with coincident events between LHAASO and ENDA.

Current simulations on ENDA indicate that it is capable of measuring the energy spectrum of cosmic rays in the knee region with detailed composition analysis. It, when added into the LHAASO, will enable LHAASO to fulfill full secondary particle measurements on the EAS with remarkable capabilities in composition separation and energy spectrum determination, which will help to obtain accurate measurement on the composition and energy spectrum in the knee region of cosmic rays.

\section*{Acknowledgments}
This work was supported in China by National Natural Science Foundation of China (NSFC, No. 12320101005, No. 12373105,  No. U2031103, No. 12205244, No. 11963004) and Hebei Natural Science Foundation (No. A2019207004).


\bibliography{library}


\end{document}